\documentclass[sigconf]{acmart}

\AtBeginDocument{%
  }

\copyrightyear{2026}
\acmYear{2026}
\setcopyright{cc}
\setcctype{by}
\acmConference[CHI '26]{Proceedings of the 2026 CHI Conference on Human Factors in Computing Systems}{April 13--17, 2026}{Barcelona, Spain}
\acmBooktitle{Proceedings of the 2026 CHI Conference on Human Factors in Computing Systems (CHI '26), April 13--17, 2026, Barcelona, Spain}
\acmPrice{}
\acmDOI{10.1145/3772318.3791054}
\acmISBN{979-8-4007-2278-3/2026/04}

\usepackage[bottom]{footmisc}
\usepackage{colortbl}
\usepackage[font=small,skip=1pt]{subcaption}
\usepackage{soul}
\usepackage{cleveref}
\usepackage{multirow}
\usepackage{graphicx}
\usepackage{cleveref}
\usepackage{tabularx}
\usepackage{xcolor}
\usepackage{caption}
\usepackage{url}
\usepackage{tabularx}
\usepackage{bbding}
\usepackage{pifont}
\newcommand{\cmark}{\ding{51}}%
\newcommand{\xmark}{\ding{55}}%

\definecolor{lightGrey}{RGB}{240,240,240}

\definecolor{revisedtext}{RGB}{0, 0, 0}
\newcommand{\revised}[1]{\textcolor{revisedtext}{#1}}

\newcommand{\dataunit}[1]{\textbf{\textsc{#1}}}

\def\controloptions#1{\underline{#1}}

\begin{document}

\title[Privacy Control in Conversational LLM Platforms] {Privacy Control in Conversational LLM Platforms: A Walkthrough Study}

\author{Zhuoyang Li}
\affiliation{%
 \department{Department of Industrial Design}
  \institution{Eindhoven University of Technology}
  \city{Eindhoven}
  \state{Noord-Brabant}
  \country{Netherlands}}
\email{z.li7@tue.nl}

\author{Yanlai Wu}
\affiliation{%
 \department{}
  \institution{University of Central Florida}
  \city{Orlando}
  \state{Florida}
  \country{USA}}
\email{wuyanlai@gmail.com}
\authornote{The author participated in this work from January to June 2025.}

\author{Yao Li}
\affiliation{%
 \department{}
  \institution{University of Central Florida}
  \city{Orlando}
  \state{Florida}
  \country{USA}}
\email{yao.li@ucf.edu}

\author{Xinning Gui}
\affiliation{%
 \department{The College of Information Science and Technology}
  \institution{The Pennsylvania State University}
  \city{University Park}
  \state{Pennsylvania}
  \country{USA}}
\email{xinninggui@psu.edu}

\author{Yuhan Luo}
\affiliation{%
 \department{Department of Computer Science}
  \institution{City University of Hong Kong}
  \city{Hong Kong}
  \country{China}}
\email{yuhanluo@cityu.edu.hk}
\authornote{Corresponding author.}


\begin{abstract}
Large language models (LLMs) are increasingly integrated into daily life through conversational interfaces, processing user data via natural language inputs and exhibiting advanced reasoning capabilities, which raises new concerns about user control over privacy. While much research has focused on potential privacy risks, less attention has been paid to the data control mechanisms these platforms provide. This study examines six conversational LLM platforms, analyzing how they define and implement features for users to access, edit, delete, and share data. Our analysis reveals an emerging paradigm of data control in conversational LLM platforms, where user data is generated and derived through interaction itself, natural language enables flexible yet often ambiguous control, and multi-user interactions with shared data raise questions of co-ownership and governance. Based on these findings, we offer practical insights for platform developers, policymakers, and researchers to design more effective and usable privacy controls in LLM-powered conversational interactions.
\end{abstract}



\maketitle


\section{Introduction}

As large language models (LLMs) become integrated into daily life---from productivity tools to emotional support and public health interventions---they gather a variety of data from our everyday activities and thus introduce new privacy challenges~\cite{zhao2024llmappstoreanalysis}.
Unlike traditional platforms that collect user information in isolated and structured data fields, \revised{conversational} LLMs \revised{process free-form user inputs via} natural languages, \revised{which can include various types of sensitive data such as} personally identifiable information, proprietary data, financial records, and medical histories \cite{Zhang2024FairGame}.
\revised{Recent work has suggested that LLMs built on decode-only Transformer (e.g., ChatGPT, Gemini) are injective, which means that a user's exact text input can be reconstructed from the output~\cite{nikolaou2025language}.} 
\revised{In other words, the sensitive personal data sent to LLMs can be regenerated} verbatim, \revised{either intentionally or unintentionally, leading to direct} privacy breaches~\revised{\cite{Yao2024survey, Zhan2024Multiuser}}.
Beyond direct data retention, LLMs can infer additional information from user input, such as the preferences and sentiment behind their language use, even when not explicitly shared~\cite{staab2023inference, Asthana2024Inference, Weidinger2022Taxonomy}, \revised{introducing additional privacy risks}. 

\revised{At the same time}, conversational interactions \revised{that involve a} virtual agent \revised{engaging} with users to gather, process, and generate information are now the dominant mode on LLM platforms~\cite{Sun2024Building, dam2024completesurveyllmbasedai}. Their human-like style can encourage \revised{more} sensitive \revised{disclosure~\cite{li2024human, Jo2024Disclosure}. While this enables the system to gather} richer responses, \revised{it} also \revised{amplifies} privacy concerns~\cite{kim2012anthropomorphism, Weidinger2022Taxonomy, Ma2025Privacy, Zhang2024FairGame}. 
\revised{Moreover, new interaction features such as model customization and conversation sharing complicate how users can control and manage their own data~\cite{Ma2025Privacy}.} 

Within previous work that investigated the privacy implications of conversational LLMs, researchers primarily focused on identifying the conversation scenarios of sensitive disclosure (e.g., the type of sensitive data)~\cite{Zhang2024FairGame}, examining users' privacy perceptions and mental models regarding emerging features on conversational LLM platforms~\cite{Zhang2024FairGame, Ma2025Privacy} and highlighting the potential privacy risks associated with conversational LLMs~\cite{WU2024ChatGPT,khowaja2024chatgpt, staab2023inference, iqbal2024llm, wu2024newerallmsecurity, Klemmer2024Using}. 

However, there remains a gap in understanding how conversational LLM platforms currently present their privacy policies and controls for users to manage how their data is being stored, trained, used, and shared. 
This understanding is crucial because control operations such as accessing, editing, retrieving, archiving, sharing, and deleting data are deeply intertwined with privacy matters, since they directly govern how personal data is handled, protected, and managed throughout its life cycle. 
\revised{Although technical implementations such as encryption and anonymization mechanisms are important, the interface ultimately determines whether users can recognize and make use of available privacy protections~\cite{Feng2021Design, Im2023Less, Lee2024PriviAware}. 
For this reason, we set out to examine the control options presented at the interface level: What end users can directly see and do.
Understanding these interface-level options can guide the design of more user-centered privacy mechanisms, which aligns with broader calls on usable privacy (e.g., General Data Protection Regulation (GDPR) Art.~12 and 25~\cite{regulation2016general}; California Consumer Privacy Act of 2018 (CCPA) code 1798.130~\cite{CCPA2018}).} In this light, we ask: \textbf{\revised{How} do conversational LLM platforms govern data practices and provide controls for users to manage their data?}

To answer the research question (RQ), we conducted an application walkthrough~\cite{light2018walkthrough} of six widely used consumer-facing conversational LLM platforms: Character.ai, ChatGPT, Claude, Gemini, Meta AI, and Pi. These platforms were identified through a \revised{three-stage} screening process guided by predefined criteria, including the presence of a consumer-facing conversational user interface (CUI), general-purpose use, popularity, \revised{and built by organizations headquartered in the U.S.} Following the application walkthrough method (an expert-driven approach focusing on examining the technical features and intended functions of a system~\cite{light2018walkthrough}),~\revised{we examined how privacy control mechanisms are governed, presented, structured, and made accessible to users. While this method does not incorporate user perspectives directly, it provides an empirical foundation for documenting current design practices and identifying areas where future user studies may be particularly valuable.}

Our findings revealed that each platform offers distinct privacy control mechanisms concerning whether, what, who, and how user data can be accessed, edited, deleted, or shared. \revised{Specifically}, we identified unique characteristics of data units, control options, and control execution mechanisms on these platforms that differ from \revised{the ones found on} conventional platforms. \revised{Given the rapid and ongoing evolution of LLM platforms, the specific interface features and control mechanisms continue to change. Nevertheless, the patterns identified in this study capture an emerging paradigm in which platforms are experimenting with new ways to define data units, explore natural language-based controls, and negotiate multi-user data sharing. These trends indicate broader directions in how privacy management is being imagined in the age of conversational LLMs.} Based on these insights, we propose \revised{empirically grounded} implications to support platform developers, policymakers, and researchers in improving data practices and identifying opportunities to strengthen user data management and privacy protection. \revised{We also outline directions for future user-centered research building on these findings.}





\section{Related Work}
Here, we first cover the related work on existing privacy control mechanisms on conventional platforms, and then describe privacy risks on conversational LLM platforms identified by researchers.

\subsection{In-platform Privacy Controls}
Privacy is \revised{a concept deeply shaped by its roots in philosophical, legal, sociological, political and economic traditions~\cite{nissim2018privacy}. In the digital age, it is typically} known as the right to control personal data, determining when, how, and to what extent it is shared~\cite{westin1968privacy}. \revised{This conceptualization of privacy focuses on informational control, which is also directly operationalized in modern regulatory frameworks. For instance, the EU's} General Data Protection Regulation (GDPR) grants users specific rights over their data, including the rights to access, rectify, erase, and restrict the processing of their information, as well as rights to data portability and to object to certain types of processing \cite{regulation2016general}.
\revised{Similarly, the California Consumer Privacy Act of 2018 (CCPA) provides rights to know, delete, and opt-out of the sale of personal information~\cite{CCPA2018}. These frameworks, despite their differences in geographical scope and legislative detail, share a common principle: empowering individuals with control over their personal data.}
For example, accessing ensures users can verify how their information is stored and used, and editing allows users to correct errors or limit oversharing. 
\revised{From users' perspectives, one common way to understand how their data is handled and to control their data is to refer to the privacy policies or} privacy notices that explain a platform's data practices~\cite{Florian2015Notices, Ebert2021Bolder, Utz2019Uninformed}. 
However, such information is often criticized for being long, vague, complex, misleading, and difficult to read, making \revised{it} easy to be ignored and ineffective in truly helping users manage their privacy~\cite{reidenberg2015disagreeable, Pollach2007policy, Salgado2023Six, Florian2015Notices, Utz2019Uninformed}.

Besides policy statements, many platforms have implemented in-platform privacy control mechanisms for users to actively manage the use of their data, including permission settings (e.g., access to \revised{built-in} sensors such as GPS and microphone)~\cite{Felt2012permissions}, information usage (e.g., third-party access) ~\cite{habib2022evaluating, Habib2021Toggles, hargittai2010facebook}, consent interfaces 
~\cite{Habib2022Cookie, Gray2021Dark}, and
more granular options for editing, deleting, or sharing their data~\cite{jacobsen2022you, schnitzler2020exploring, yilmaz2021perceptions}, etc. Extensive research has explored the design of these control features, such as the visual presentations and the timing of prompts, as well as how they influence the ways that users control their data~\cite{habib2022evaluating, Feng2021Design, Habib2019websites, Habib2021Toggles, jacobsen2022you}. Researchers noted that although these features provide users with a degree of control, there were several usability barriers that hinder users from configuring them effectively~\cite{Leon2012Opt, Habib2020Usability, Murillo2018If}. For example, Habib et al.~\cite{Habib2020Usability} found that users struggled with locating the appropriate settings page and understanding the content of opt-out controls~\cite{Habib2020Usability}. 

To address these usability barriers, prior work has proposed various solutions. For example, Liu et al. developed a ``\textit{personalized privacy assistant}'' that provides recommendations and nudges to help users stay aware of the privacy choices they have previously consented to and encourage them to review and adjust these choices~\cite{Liu2016Follow}. Other research has also explored supporting users in managing permission settings by profiling their privacy preferences~\cite{Lin2014Modeling}.
\revised{Nevertheless, these strategies primarily operate on conventional platforms where user data is collected in isolated, structured fields~\cite{habib2022evaluating, Habib2021Toggles, hargittai2010facebook}, an interaction mode that was more common before the rise of LLMs. On modern LLM platforms where conversational interfaces become the mainstream, the input of unstructured natural language creates a fundamentally different privacy dynamic, because users can no longer easily foresee what personal data is being revealed or derived when engaging in free-form conversation, as every utterance can become a data point for sensitive inference~\cite{Zhang2024FairGame}.} This shift highlights the need to investigate new privacy control paradigms tailored to conversational LLM platforms.

\subsection{Privacy in Conversational LLM Platforms}
While interactions with conversational agents (CA) powered by LLMs enabled rich and natural exchanges, researchers have highlighted the privacy risks associated with personal data disclosure in these interactions~\cite{Zhang2024FairGame, Ma2025Privacy, Asthana2024Inference}. A primary concern is ``memorization,'' which manifests in two ways.
First, LLMs store conversation history to maintain context, often retaining a multitude of user data, including personally identifiable information (e.g., names, email addresses, age), sensitive experiences (e.g., health records, finances, emotions), and thoughts, posing significant privacy risks~\cite{Zhang2024FairGame}. \revised{Despite the common belief that the black-box nature of LLMs inherently obscures user data, a recent work showed that user input to LLMs can be inverted, creating a risk of direct input recovery~\cite{shanmugarasa2025SOK}}. Second, LLMs may retain and use user-provided information for model training, improving the performance of future models but inadvertently increasing the risk of privacy breaches, as these models can unintentionally leak memorized user information in responses to others~\cite{yang2024memorization, Zhang2024FairGame, aditya2024evaluating, lukas2023analyzing, carlini2021extracting}. For instance, Zhang et al. documented a memorization leak where a participant using~\revised{an LLM plugin for programming software} experienced an unexpected disclosure: \revised{Typing} a classmate’s name triggered an auto-completion suggestion revealing their school ID~\cite{Zhang2024FairGame}.

Furthermore, as CAs on LLM platforms were often designed with contextual understanding and empathy, many users perceive them as companions and willingly share private and sensitive information, even though the platforms were not designed for such disclosures~\cite{Jo2024Disclosure, kim2012anthropomorphism, li2024human, Weidinger2022Taxonomy}. This risk can be further amplified with LLMs' ``reasoning'' ability, which can derive personal information from users' inputs, potentially revealing details users did not explicitly disclose~\cite{staab2023inference, tomekcce2024private, yukhymenko2024synthetic, li2024human, Asthana2024Inference}. For instance, Staab et al. demonstrated that LLMs can infer personal location, income, and gender based on textual patterns in user-provided content~\cite{staab2023inference}. These findings heightened the privacy risks, as users may unintentionally disclose more information than they realize~\cite{staab2023inference}.

As LLMs are increasingly integrated into various platforms to support our daily work and life~\cite{tao2024harnessing}, empowering users to control their privacy is more critical than ever. Despite knowing the potential privacy risks and users' concerns, there is a lack of \revised{an} overview of how platforms implement data control mechanisms or how effective these mechanisms might be. To address this gap, we walked through data control features and mechanisms on six \revised{widely used} conversational LLM platforms, marking a first step toward designing more effective privacy control support and establishing privacy guidelines for human-LLM interactions.



\section{Method}
To \revised{examine} data control features provided by existing conversational LLM platforms and the mechanisms behind these features, we conducted an application walkthrough~\cite{light2018walkthrough} covering six consumer-facing platforms from November 19th 2024 to January 2nd 2025.
\revised{In the following, we first explain what an application walkthrough method is and the rationale of choosing this method. We then present} how we \revised{searched and} identified \revised{target} platforms, conducted the walkthrough, and performed data analysis.

\subsection{\revised{Methodological Considerations}}

\subsubsection{\revised{Application Walkthrough as an Expert-driven Approach}}
\revised{Application walkthrough\footnote{\revised{The application walkthrough method employed in this study differs from the ``cognitive walkthrough'' method, which evaluates usability by examining how a first-time or infrequent user would carry out predefined \textbf{tasks}~\cite{Mahatody2010Cognitive}. In contrast, our walkthrough follows Light et al.’s~\cite{light2018walkthrough} approach, focusing on how platforms structure and present data control \textbf{features}, rather than on assessing usability.}} is a} widely adopted approach for researchers to explore the features and functionalities of digital interfaces, websites, and software applications through a socio-technical lens~\cite{ONeill2024Food, Moradzadeh2024fake, Reddy2024BeReal, Reime2023Walking}. 
\revised{This method is} often led by domain experts, as they possess the necessary knowledge to identify potential usability issues and evaluate the effectiveness of data control mechanisms.
Similar expert-driven inspection methods (e.g., heuristic evaluation, cognitive walkthrough) have also been widely applied in usable security and privacy research~\cite{AbuSalma2020Evaluating, smith2020can, whitten1999johnny, habib2022evaluating, dieter2020patterning, Malki2024mHealth}, \revised{which can yield empirical contributions by generating structured observations of platform features and design patterns~\cite{Wobbrock2016Contribution}.} 

\revised{We chose} application walkthrough \revised{to answer our RQ, because it} enabled us to document exposed settings, fine-grained controls, and nuanced privacy terminology---elements that are often hidden, technically complex, or require expert interpretation to fully understand~\cite{light2018walkthrough}. 
\revised{Therefore, an expert-driven walkthrough enables researchers to identify potential gaps or design opportunities by examining how interface features are presented to users. As suggested by Light et al.~\cite{light2018walkthrough}, such analytic groundwork is crucial for building a concrete understanding of current design trends and informing future user-centered investigations.}

\subsubsection{\revised{Analytic Lens Within Temporal Limits}}~\label{analytic-lens-temporal}
\revised {While conducting the walkthrough, we were aware that} the interface designs and control features on LLM platforms were evolving rapidly, and \revised{since then,} new models such as Grok4~\cite{grok2025} have emerged. \revised{We are also aware that such walkthroughs often require in-depth analyses across several months. It is therefore natural that an analysis of multiple platforms spans over a long time, during which incremental updates inevitably occur~\cite{Habib2020Usability}.
Hence, we acknowledge that the walkthrough captures only a snapshot describing emerging design paradigms rather than an exhaustive, continuously updated catalog of all conversational LLM interfaces.} 

\revised{However, this temporal scope does not compromise the validity of our findings. Our goal was not to document every interface feature that updates over time, but to identify the underlying interaction paradigms through which platforms conceptualize and operationalize user data control, which persist even as individual interface elements evolve. By examining these broader design paradigms instead of version-specific features, the walkthrough uncovered design directions that continue to characterize contemporary conversational LLM platforms.}


\subsubsection{\revised{Researchers' Positionality}}
\revised{Given that application walkthrough is an expert-driven and time-bound approach, here we disclose our positionality and expertise for transparency~\cite{holmes2020researcher, Sousa2025Positionality}.} 
All authors are experienced qualitative researchers in human-computer interaction and ``early adopters'' of the emerging features brought by conversational LLM platforms in our daily lives. In particular, the first and the corresponding authors have conducted several research projects on conversational LLM platforms; the other authors have several years of research experience in usable privacy and security, making the team well-equipped to conduct this application walkthrough and data analysis.


\begin{figure}[ht]
         \centering
         \includegraphics[width=0.48\textwidth]{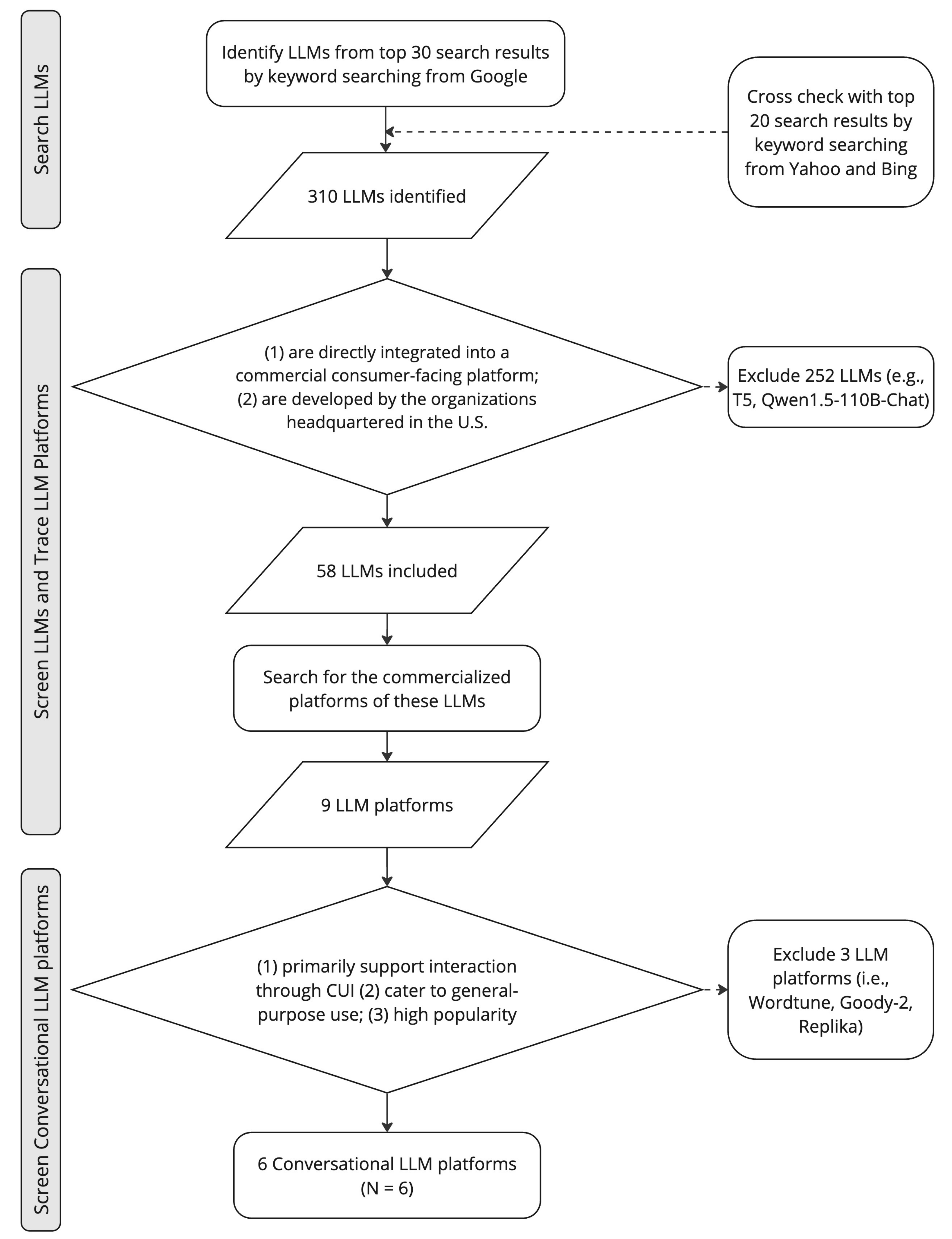}
         \Description{The three-stage data collection process.}
          \caption{\revised{The three-stage data collection process.}}
          \label{fig:DataCollection}
\end{figure}


\subsection{Platform Identification and Selection}
Our data collection process consisted of three stages (see Figure~\ref{fig:DataCollection}): (1) a search and screening of LLMs across three search engines; (2) tracing the models back to the consumer-facing platforms where the models are deployed and the user interaction data are directly managed by the model developers; and (3) screening conversational platforms to meet our inclusion criteria, which are detailed in the following section. 

\subsubsection{Searching LLMs}~\label{searchLLMs}
First, we identified a set of LLMs using the three most widely used search engines in the United States as of June 2024: Google (87.14\% market share), Bing (7.81\%), and Yahoo (2.58\%)~\cite{Statcounter2024Engines}. Before conducting the searches, we cleared the search history and enabled incognito mode in the Chrome browser~\revised{and set the device’s IP location to the United States to minimize location bias in search}. We also set the time range for search results to between November 2022 (when LLMs gained significant attention following the launch of ChatGPT~\cite{ChatGPT} in November 2022) and June 2024 (the month we conducted our searches). Our search process focused on \revised{identifying unique language models (e.g., GPT-3.5-Turbo), rather than platforms that use existing models}. The search query was:

~\textit{(``Large language models'' OR ``LLMs'' OR ``Generative AI'' OR ``GenAI'' OR ``Chatbot'' OR ``Conversational AI'') AND (``overview'' OR ``list'' OR ``survey'' OR ``review'' OR ``popular'' OR ``best'' OR ``top'' OR ``most used'')} 

\revised{Given Google's dominant market share (87.14\%)~\cite{Statcounter2024Engines}, we started from Google to look into the top 30 search results (highest-ranked links returned for the search query while excluding advertisements and sponsored content to maintain the objectivity and reliability of our data collection)} and generated an initial list of LLMs mentioned. \revised{Then, we} cross-checked the top 20 search results from Bing and Yahoo, to complement our list; during that process, we did not find any new LLMs beyond what had already been identified in the Google search results, \revised{indicating that data saturation had been reached~\cite{fusch2015we}; we therefore concluded our search at that point}.

The search results included academic papers, news articles, industry reports, leaderboards (i.e., systematic summaries of existing LLMs on open-source platforms such as Chatbot Arena~\cite{ChatbotArena}), and product reviews. 
The first author reviewed these results and extracted 310 LLMs, including both the latest versions of these language models and their historical iterations.


\begin{table*}
 \def\arraystretch{1.1}
 \small
 \caption{Conversational LLM platforms that are analyzed in this study and the time frames in which we conducted the application walkthrough.}
 \label{tab:LLMPlatforms}
 \begin{tabularx}{\textwidth}{ | p{1.3cm} | p{1.3cm} | p{3.0cm} | X | p{1.8cm} |} 
 \rowcolor{lightGrey}
 \hline
 \textbf{Platform} & \textbf{Developer} & \textbf{Employed Models (Oct 2024)} & \textbf{Popularity} & \textbf{Walkthrough Time Frame}\\ 
 \hline

Character.ai & Character.ai & In-house model & ``\textit{Character.ai has over 20 million monthly active users in 2024.}''~\cite{Popularity_CharacterAI} & Dec 4--10, 2024\\ \hline

ChatGPT	 & 	Open AI	 & 	GPT-3.5, GPT-4, GPT-4o, GPT-4o mini, DALL·E & 	``\textit{ChatGPT currently has over 180 million users.}''~\cite{Popularity_ChatGPT} & Nov 19--26, 2024 \\ \hline

Claude & Anthropic & Claude 3.5 Sonnet, Claude 3 Sonnet, Claude 3 Opus, Claude 3 Haiku & ``\textit{The website sees nearly 54.4 million visitors every month.}''~\cite{Popularity_Claude} & Dec 25--30, 2024\\ \hline

Gemini	 & 	Google	 & 	Gemini 1.5 Flash, Gemini 1.5 pro & ``\textit{Google Gemini has an average of 274.7 million monthly visits by September 2024.}''~\cite{Popularity_Gemini} & Dec 26, 2024--Jan 2, 2025\\ \hline

Meta AI & Meta  &  Llama 3.1- 405B	 & ``\textit{The assistant(Meta AI) reached 400 million monthly active users and 40 million daily active users in early August.}''~\cite{Popularity_MetaAI} & Dec 23--26, 2024\\ \hline

Pi	 & 	Pi	 & 	Inflection-2.5	 & ``\textit{Our one million daily and six million monthly active users have now exchanged more than four billion messages.}''~\cite{Popularity_Pi} & Dec 19--23, 2024 \\ \hline

\end{tabularx}
\end{table*}


\subsubsection{Screening LLMs and Tracing LLM Platforms}
In this step, we traced the platforms that originally built and deployed these LLMs (e.g., ChatGPT~\cite{ChatGPT}) rather than third-party platforms that integrate LLMs via APIs (e.g., Poe~\cite{Poe}). This is because the latter platforms lack full control over the model behavior, and thus are not directly accountable for the privacy risks, such as potential data breaches. Additionally, they often host multiple LLMs from different developers, introducing complexity that warrants further research.
In addition, while LLMs are generally trained for multilingual tasks, English remains the dominant source of their training data~\cite{guo2024LLMEnglish}. \revised{Additionally, we focus on models developed by organizations headquartered in the U.S., because data practices of digital platforms are closely tied to regional regulatory and commercial contexts~\cite{lim2025navigating}. Exploring models built in other cultural and regulatory environments, such as Ernie, which must comply with China’s Personal Information Protection Law (PIPL)~\cite{PIPL2021}, would require a broader cross-cultural and cross-regulatory analysis that is beyond the scope of this paper~\cite{Ghaiumy2021Culture, li2022cultural, Cho2018Cultural}. We point to Section~\ref{LimitationFutureWork} for how future work can take up this comparative perspective.} 

\revised{Based on these considerations,} we further screened the 310 LLMs identified in the previous search results based on the following criteria:

\begin{itemize} 
    \item Models should be deployed by a commercialized consumer-facing platform, where the user interaction data are directly managed by the model developers.
    \item Models should~\revised{be developed by organizations headquartered in the U.S., meaning their primary regulatory obligations fall under U.S. frameworks. As a result,} their default interface language is English, their official documentation and policies are primarily in English, and their developers provide support in English first.
\end{itemize}

During this process, we excluded 196 models that are not commercialized (e.g., people can only use T5~\footnote{T5 is an LLM introduced by Google Research in 2019.} by installing the API, but it is \revised{neither} commercialized\revised{, nor} designed for non-expert consumers). We also excluded 47 models because they are not developed by the organizations headquartered in the U.S. (e.g., ERNIE 3.0 Titan is developed by Baidu, headquartered in China; Mistral-7B is developed by Mistral AI, headquartered in France).
For the remaining 58 models, we traced back to the developers of these models and their initial conversational interfaces, leading to 9 platforms\revised{, namely, Character.AI, ChatGPT, Claude, Gemini, Meta AI, Pi, Replika, Goody2, Wordtune}.

\subsubsection{Screening Conversational LLM Platforms}
With the platforms identified in the previous step, we conducted further screening based on the following criteria: 

\begin{itemize}
    \item The platform should primarily support interaction through conversations, featuring a clear user interface that includes elements such as a user input field and LLM-generated responses displayed in chat bubbles (e.g., Wordtune~\cite{wordtune}~\revised{is excluded, because} it is an LLM-powered browser extension for reading and writing, and its users can receive rewrite suggestions by highlighting specific text,~\revised{but it does not have a conversational interface}). 
    \item The platform should explicitly state to provide general-purpose support across various domains such as productivity, journaling, and entertainment, rather than focusing on a specific domain. For example, Replika~\cite{replika} was excluded because it is advertised as ``\textit{an empathetic friend}'' in a 3D anime-style environment, positioning it primarily as an application focused on emotional support. This positioning aligns with the use scenarios identified in prior studies~\cite{possati2023psychoanalyzing, depounti2023ideal, laestadius2024too}.\revised{By contrast, Character.ai was included because, despite its emphasis on character-based interactions, it allows users to create or engage with characters for a wide range of purposes}. For example, there are some categories provided in the navigation bar, such as ``\textit{assistants},'' ``\textit{anime},'' ``\textit{learning},'' ``\textit{lifestyle},'' etc.
    This criterion ensures that our analysis captures conversational LLM platforms designed for diverse user interactions rather than niche applications.~\label{screening-purpose}
    \item The platform is widely used with validated sources reporting its user base (e.g., Goody-2~\cite{Goody2} was excluded because we could not find validated information about its user base by the time of data collection). This criterion ensures that our walkthrough focuses on platforms with large real-world impact.
\end{itemize}

These criteria ensure that our findings apply to a broad range of LLM interactions rather than being restricted to specific use cases. They also help focus our analysis on platforms with real-world impact and widely adopted privacy control mechanisms.
As such, we excluded 3 platforms and included 6 platforms for the final data analysis, \revised{namely Character.ai~\cite{Characterai},
ChatGPT~\cite{ChatGPT},
Claude~\cite{Claude}, 
Gemini~\cite{Gemini}, 
Meta AI~\cite{MetaAI}, 
and Pi~\cite{Pi}}, see Table~\ref{tab:LLMPlatforms} for more details. 
\revised{We} only focused on the versions for individual use, rather than enterprise use, API integration, etc.

\subsection{Walkthrough Procedure}~\label{walkthrough}

We followed Light et al.'s application walkthrough method, which involves researchers navigating through the interface or system in a structured manner while documenting their observations and experience, taking notes of interface elements, navigation flows, content organization, and usability~\cite{light2018walkthrough}. 
\revised{Specifically, application walkthrough includes two steps: analyzing the \textit{expected environment of use} and \textit{technical walkthrough}, which we detail below.}

\subsubsection{\revised{Analysis of Expected Environment of Use}}
    \revised{This part refers to each} platform's~\textbf{vision} (i.e., the anticipated usage scenario and target users),~\textbf{operating model} (i.e., business strategies, such as premium versions or in-app purchases), and~\textbf{governance} (i.e., the strategies employed by the service provider to oversee and regulate user activity to support their operating model and fulfill their vision). \revised{Thus, we analyzed} relevant institutional materials encountered during our navigation through the platforms, such as Terms of Service, Privacy Policies, FAQs, and other relevant documents. 

\subsubsection{\revised{Technical Walkthrough}}
\revised{We} examined the mechanisms that allow users to control their personal data, such as opt-in/out, accessing, editing, updating, and deleting personal information. We also considered how these options can be executed by users (e.g., indicating their choice by using toggles), which is key to privacy control. 
To document any variations in data control options available to users based on their account status and subscription plan, we examined all the platforms as a signed-out user, a signed-in unpaid user, and a signed-in paid user, respectively. For features involving sharing among multiple users (e.g., ChatGPT allows users to share their conversation with other users through an auto-generated shared link), we used a second user account to examine the platform from the perspective of a user who receives shared content. The time frames of walking through each platform are specified in Table~\ref{tab:LLMPlatforms}. 

\revised{\textbf{Initial Walkthrough Protocol.}}
\revised{To develop a consistent walkthrough protocol across platforms, we began with an initial walkthrough of ChatGPT and Character.ai, because the former was the most widely used platform at the time of data collection, while the latter adopts a distinct vision and operational model. Examining both allowed us to cover different design considerations and interaction paradigms at the outset. The first author meticulously documented every interface deemed to contain data control features (e.g., pop-up windows, page refreshes) during the initial interaction with ChatGPT, resulting in 95 distinct screenshots accompanied by detailed field notes. These notes captured not only the interface elements but also contextual information about the interaction (e.g., whether the feature was accessed while signed in or signed out, and the interaction flow leading to the interface). Each screenshot could contain multiple data control options; for example, a single drop-down menu might simultaneously display edit, delete, and share functions.}

\revised{After this, all authors reviewed the screenshots and notes together to determine which parts of the interface required further examination. For instance, in the initial walkthrough, the first author experimented with different types of personal information to see whether they would trigger ChatGPT’s memory feature. However, through our discussions, we realized that the analysis should focus only on interface-level data control options, rather than backend or undocumented mechanisms, because such implementations cannot be reliably verified through a walkthrough method. As a result, any screenshots or notes outside this scope were excluded from further analysis.}

\begin{figure*}
    \centering    
\includegraphics[width=0.8\textwidth]{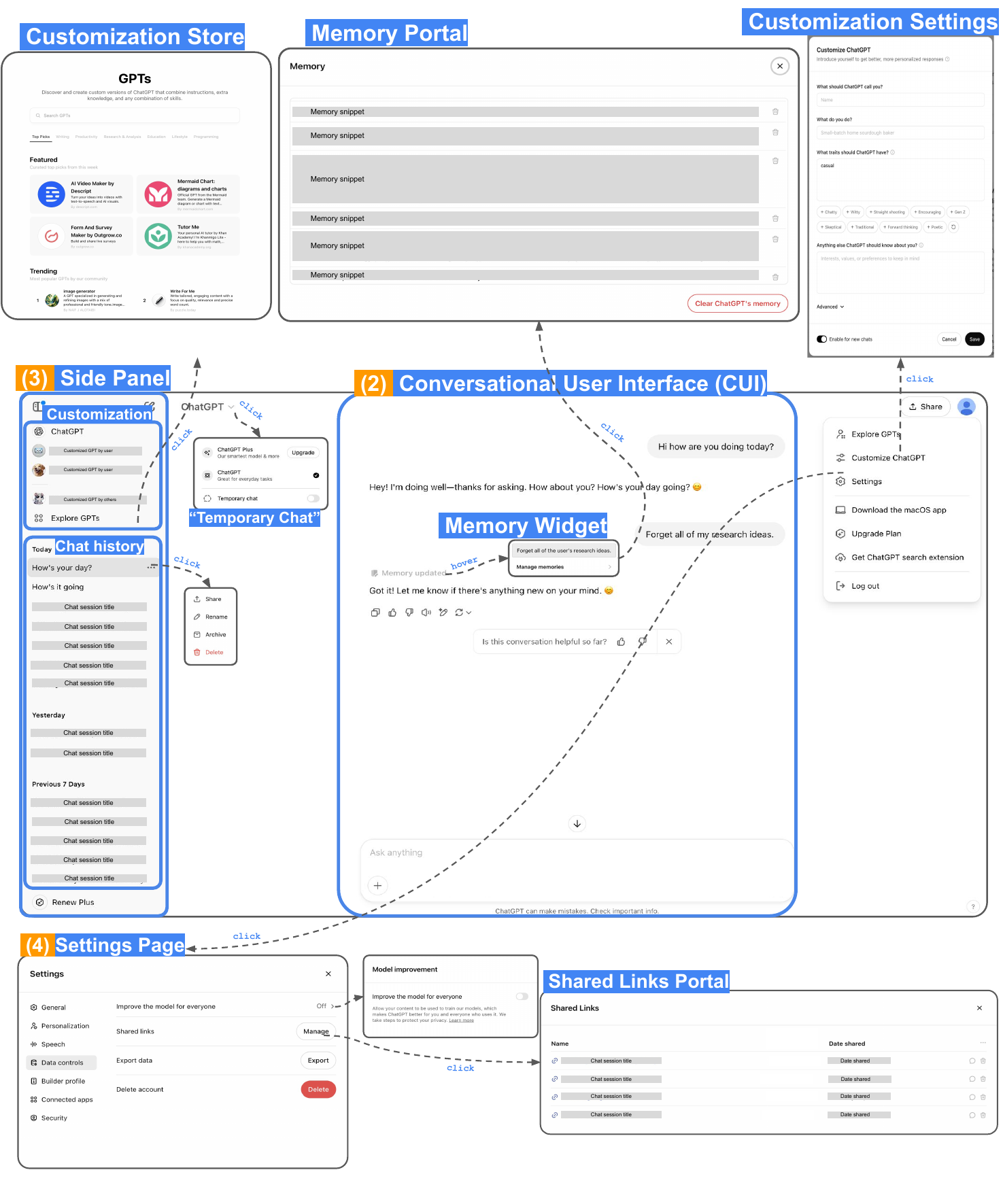}
    \Description{Interface examples of ChatGPT (free plan).}
    \caption{\revised{Interface examples of ChatGPT (free plan).
    \textbf{Conversational User Interface (CUI)} is the main interface where users interact with ChatGPT.
    \textbf{Memory Widget} appears when a user input triggers the memory-related feature, where users can view memory snippets and access the memory portal. 
    \textbf{Side Panel} is typically located next to the chat window, which provides access to:
    \textbf{chat history}, organized by chat sessions; 
    \textbf{customization} GPTs created by other users and Customized GPTs created by the user;
    \textbf{customization Store} for discovering and interacting with customized CAs created by other users.
    \textbf{Settings Page} is a centralized hub for managing various user preferences.
    \textbf{Shared Links Portal} provides access to and control on shared conversations. 
    \textbf{Customization Settings} allows users to input descriptive information to customize their ChatGPT.
    \textbf{Memory Portal} is a space where users can manage all stored memory snippets.}} 
    \label{app:interfaces}
\label{fig:InterfaceANDflow}
\end{figure*}

\revised{The first author then conducted a second technical walkthrough of Character.ai, producing 49 screenshots with notes. Next, all authors jointly reviewed the materials and identified the relevant features for the study. These discussions led us to develop a walkthrough protocol} (\revised{the interface terminology used throughout the paper is illustrated in Figure~\ref{fig:InterfaceANDflow}}, using ChatGPT as an example). The final overall technical walkthrough flow~\revised{and the specific steps are outlined chronologically}:
\begin{enumerate}
    \item Boarding page:~\revised{We} reviewed the institutional materials, such as Privacy Policy, Terms of Use, and the linked materials.
    \item Conversational User Interface (CUI): ~\revised{We} initiated several rounds of conversation using topics potentially containing personal information (e.g., \revised{``\textit{My name is Johnny. I am an HCI researcher. My research is about usable privacy.}''; ``\textit{I live in city X and I want some trip advice for city Y.}''; and ``\textit{I had two meetings today. One is with my colleague A, and another is with my supervisor B. I now want to improve my presentation skills.}''). Their primary function was to serve as consistent stimuli for eliciting relevant interface elements across platforms, rather than to compare how different inputs were processed or to examine different conversational topics. We therefore used short, neutral conversations and made minor adjustments when necessary to trigger comparable data-control options across systems (e.g., we added ``\textit{Remember...}'' to trigger memory-related features).}
    During these conversations, we explored all the available control operations (e.g., sharing the conversation, regenerating response) and linked interfaces (e.g., Memory Widget, Memory Portal) that emerged in the CUI.
    \item Side panel (Chat history):~\revised{We} accessed, edited and deleted chat history. \revised{We} also interacted with other control options (e.g., archive, rename the session).
    \item Side panel (Customization):~\revised{We} interacted with customization-related features (e.g., \revised{customize a conversational agent to be ``\textit{coach that can improve my presentation skills} or customize a ``Project'' for ``giving feedback on a walkthrough research project'' by uploading some prior literature). We also tried to} access, edit, delete, and other linked controls (e.g., share the customized persona) on the customization records and customize items through all linked interfaces (e.g., Customization Store). 
    \item Settings page:~\revised{We} performed all the control options available (e.g., access shared message, export data, opt-out on model training) and explored all the linked interfaces (e.g., Shared Links Portal).
    \item Other control options:~\revised{We} interacted with available options related to data control (e.g., Turn on the ``\textit{Temporary Chat}'' mode, view the ``\textit{Customization settings}''). 
    \item \revised{Log out: We} deleted all the data and logged out.
\end{enumerate}


When testing control options that would influence the conversational interactions, we always returned to the CUI to initiate new conversations. For example, after customizing a CA persona, we started a new conversation with the customized CA and explored operations such as accessing, editing, deleting, and sharing the chat history. Similarly, after sharing a conversation, we used the second user account to continue the conversation and examine what control features were enabled for the receiving user. 

\revised{\textbf{Protocol Refinement.}}
While the technical walkthrough structure remained consistent across platforms, slight adjustments were made to accommodate each platform's unique layout and feature set. For instance, on Character.ai, the CUI is flanked by two side panels: \revised{The} left panel provides access to previously interacted ``characters,'' option to create new ones, and settings, while the right panel displays details about the interacting character, including chat history with that character, pinned messages, and options for voice and style customization. We therefore conducted the technical walkthrough on Character.ai by the sequence: boarding page, CUI, right side panel, left side panel, settings page, \revised{and} other control options. 

\revised{Using this refined protocol, we retrospectively checked screenshots and field notes from ChatGPT and Character.ai, and discarded those that are not relevant. We then walked through the remaining platforms, collecting screenshots and writing field notes for Claude (38 screenshots), Gemini (33 screenshots), Meta AI (17 screenshots), and Pi (13 screenshots). The smaller number of screenshots for Meta AI and Pi reflects their limited data-control features, as they did not have customization and memory-related functions, and their sharing mechanisms were relatively simple. All the walkthrough logs are documented in an Excel file, with columns ``\textit{Platform Name},'' ``\textit{Derive time},''``\textit{Links},'' ``\textit{Screenshots},'' ``\textit{Field Notes},'' and ``\textit{Initial Codes},'' see Appendix~\ref{app: analysis} for some examples of our walkthrough logs.} 
For presentation purposes, the screenshots included in this paper were post-processed with adjustments to color, contrast, and lightness to enhance readability, and the information potentially disclosing the authors’ identities was redacted.



\begin{table*}
 \footnotesize
 \centering
 \caption{Adapted data practices scheme.}
 \label{tab:PrivacyScheme}

 \begin{tabularx}{\textwidth}{| c | X |} 
 \hline
 \rowcolor{lightGrey}
 \textbf{Category} & \textbf{Description} \\ 
 \hline

Data Ownership & Who is responsible for and fully controls the information, and what information do they own~\cite{Branscomb1994Owns, asswad2021data}.
\\ \hline

First Party Collection/Use & Whether, how and why a service provider collects and uses information. 
\\ \hline

Third Party Sharing/Collection & Whether, how and why user information may be shared with or collected by the third party. 
\\ \hline

\revised{User Choice} & \revised{Whether and how users can make choices regarding choice type, choice scope, personal information type, purpose, and user type.} 
\\ \hline

User Access, Edit, and Deletion & Whether and how users may access, edit, or delete their \textit{personal data}. 
\\ \hline

Data Retention & How long user information is stored. 
\\ \hline

Data Security & How user information is protected. 
\\ \hline

Policy Change & Whether and how users will be informed about changes to the privacy policy.
\\ \hline

Do Not Track & Whether and how Do Not Track signals for online tracking and advertising are honored. 
\\ \hline

* Data sharing among multiple users & What and how can information be shared among multiple users. Whether and how users interact with and control the shared information.
\\ \hline
\end{tabularx}

\vspace{2mm}

\begin{minipage}{\textwidth}
   \small \textbf{\textit{Note}}: This scheme is primarily derived from Wilson et al.'s annotation scheme that captures the data practices included in privacy policies~\cite{Wilson2016Scheme}, and synthesized with other studies~\cite{sovern1999opting, Bui2022Optout, Branscomb1994Owns,asswad2021data}. Key adjustments include: (1) revised category wordings for clarity, for example, we changed ``User Choice/Control'' to \revised{``User Choice,'' because in our study's context, user control encompasses both user choice and user access, edit, and deletion}; (2) added ``Data Ownership'' that is not part of the original scheme, but an important prerequisite for exercising data rights~\cite{Branscomb1994Owns, asswad2021data}; (3) excluded ``International \& Specific Audiences'' from the original scheme, \revised{since incorporating multilingual or international platforms would conflate heterogeneous regulatory, cultural, and socio-technical contexts. Such cross-contextual analysis warrants a dedicated comparative framework, which we leave for future work}; (4) added new items as the walkthrough processed, marked with ``*''.
\end{minipage}

\end{table*}


\subsection{\revised{Data \&} Analysis}~\label{Method: Analysis}
Our analysis focused on two main parts of the data: information related to data governance based on the platforms’ expected environment of use, and data gathered through the technical walkthrough (see Section~\ref{walkthrough}). We focused on data governance because it is a key aspect of understanding how platforms define and manage the privacy-related data practices and user control. Data Governance refers to the regulation regulating data availability, usability, integrity, and security to ensure consistent handling in compliance with policies and regulations~\cite{khatri2010designing}. The other two aspects of the expected environment of use (i.e., vision and operating model) were clearly stated by each platform; thus, we did not conduct thematic analysis on them.

The first author began the analysis by examining ChatGPT and Character.ai, as they embody different visions and expected usage, while both platforms cover a wide range of use cases. Character.ai emphasizes user-driven customization and character sharing, whereas ChatGPT emphasizes versatility for everyday life.
\revised{All our analysis was done manually without any AI-assistance.}

\subsubsection{Governance-related information}~\label{method: governance}
For analyzing data governance, we employed a hybrid approach~\cite{fereday2006demonstrating}, combining deductive application of existing categories from the privacy policy annotation scheme developed by Wilson et al.~\cite{Wilson2016Scheme} with inductive theme development. First, the first and corresponding authors reviewed institutional materials for ChatGPT and Character.ai (e.g., Privacy Policies and Terms of Use). Next, all authors collaboratively revised the privacy policy annotation scheme to suit the study context. This scheme, originally developed to annotate a substantial dataset (i.e., ``\textit{115 privacy policies (267K words) with manual annotations for 23K fine-grained data practices}'')~\cite{Wilson2016Scheme}, has been widely adopted in usable privacy \revised{research~\cite{Barth2022Understanding, Windl2022Automating}}. We tailored the scheme to the current context by incorporating insights from relevant literature.

During analysis, we identified an emergent theme, ``data sharing among multiple users,'' in Character.ai’s Privacy Policy. Acknowledging the relevance of this theme to data governance practices and its absence in the original scheme, we incorporated it into our revised analytical framework. The adapted scheme used in the final analysis is presented in Table~\ref{tab:PrivacyScheme}. The first author then integrated governance-related codes from the remaining four platforms into this scheme. 

\subsubsection{Data from Technical Walkthrough}~\label{terms}
Our technical walkthrough started with an inductive approach to develop initial indices, followed by a deductive approach that applied these indices in subsequent analysis.
In the inductive analysis phase, the first author applied an inductive approach to analyze the field notes and screenshots from the technical walkthrough of ChatGPT and Character.ai. This process led to the generation of initial codes, such as ``\textit{Granularity on controllable information (access, edit, deletion, and sharing).}'' These preliminary codes were then discussed in meetings with all authors to synthesize insights from the technical walkthrough of these two platforms. Through this process, we observed that user data recorded on these platforms is not limited to chat history but also the memory stored about the users and their customized objects, such as descriptions of a conversational character on Character.ai. We also identified nuances in data control units (e.g., Character.ai allows users to customize both the conversational character and \revised{roles for users themselves to play on}), interaction methods to perform controls (e.g., natural language commands for managing ``memory'' in ChatGPT), and data-sharing features (e.g., users can continue conversations shared by others in ChatGPT).
Based on these preliminary findings, we developed \revised{three categories} for subsequent analysis:
\begin{itemize}
    \item \textbf{Data Types and Units:} Categories of user data recorded on the platforms (e.g., chat history, customization record, memory) and the ways they are structured as discrete \revised{elements for control (e.g., one chat session, one piece of memory snippet, one customized conversational agent)}.
    \item \textbf{Control Options:} Actions users can take on their data (e.g., access, retrieval, sharing, memorization) or choices they can make (i.e., opt-in and opt-out) over first/third-party data collection and usage.
    \item \textbf{Control Executions:} Methods and effects for executing control operations (e.g., graphical user interfaces (GUIs) in settings, natural language commands in chat).
\end{itemize}

Next, our analysis of the four remaining platforms followed a deductive approach, strategically guided by the categories derived from the walkthroughs of ChatGPT and Character.ai. \revised{To ensure consistency in the analyzing process, the first author developed a form documenting each data type and unit, the platform, and specific examples from each platform. All authors met regularly to review emerging examples and discuss ambiguous cases. For instance, after discussion, we treated Character.ai’s ``\textit{pinned message}'' as a form of \textit{memory} rather than merely a conversational message, because pinned messages can be referred to across chat sessions (see Section~\ref{Units_Memory}). Similarly, we categorized Claude’s ``\textit{Projects}'' as a type of \textit{customized object} rather than a new data type, since its primary function---like other customized objects---is to allow users to personalize their conversations (see Section~\ref{customization}). Upon finishing the analysis,} no new categories emerged beyond those established in the initial walkthroughs. \revised{The final sub-themes and themes are reflected in the sub-sections and section titles presented in the Findings. The detailed collaborative and iterative analysis process is provided in Appendix~\ref{app: analysis}.}



\section{Findings}

In this section, we first summarize the visions, operating models, and data governance of the six platforms, and then detail what data can be controlled and how these controls are executed on the platforms. First, all platforms share a common \textbf{vision} of harnessing LLMs to humanize digital interactions across a wide spectrum of life. Yet, each of them prioritizes different aspects to achieve this vision: Character.ai emphasizes interactive customization and entertainment by featuring diverse user-created characters on its homepage; ChatGPT, Gemini, and Meta AI position themselves as versatile assistants to augment creativity and productivity (e.g., the tagline of Meta AI ``\textit{Ask Meta AI anything.}''); Claude branded itself more for productivity purposes, saying ``\textit{Claude is a next generation AI assistant built by Anthropic and trained to be safe, accurate, and secure to help you do your best work};'' and Pi highlights its emotional intelligence, branding itself as ``\textit{The first emotionally intelligent AI}.''

Regarding \textbf{operating models}, Character.ai, ChatGPT, Claude, and Gemini offer subscription plans for individual use with more advanced model capabilities (e.g., allowing uploading files, faster and more messages), early access to new features, and additional features that are not available to unpaid users (e.g., ``\textit{customize GPT}'' in ChatGPT, ``\textit{Project}'' in Claude, and ``\textit{Save Info}'' in Gemini), while Meta AI and Pi do not have subscription plans. 

With the understanding of each platform's vision and operating model, we present findings from the analysis of data governance and the technical walkthrough below.


\begin{table*}[h]
\renewcommand{\arraystretch}{1.2}
\small
\centering
\caption{Data governance practices stated in the platforms' \revised{Privacy Policies or Terms of Use. 
\cmark~indicates that the platform claims to offer the listed practice, whereas \xmark~indicates that it does not.}}
\label{tab:governance}

\begin{tabular}{|p{0.16\textwidth}|p{0.18\textwidth}|p{0.09\textwidth}|p{0.08\textwidth}|p{0.07\textwidth}|p{0.07\textwidth}|p{0.08\textwidth}|p{0.07\textwidth}|}
\hline
\multicolumn{2}{|c|}{\textbf{Governance / Platform}} &
\textbf{Character.ai} & \textbf{ChatGPT} & \textbf{Claude} &
\textbf{Gemini} & \textbf{Meta AI} & \textbf{Pi} \\
\hline

\multirow{2}{*}{Data ownership}
& Ownership of user input remains with the user
& \cmark & \cmark & \cmark & \cmark & \cmark & \cmark \\
\cline{2-8}
& Ownership of model output remains with the user
& \cmark & \cmark & \cmark & \cmark & \cmark & \xmark \\
\hline

\multicolumn{2}{|l|}{User access, modification, and deletion of their data}
& \cmark & \cmark & \cmark & \cmark & \cmark & \cmark \\
\hline

\multirow{2}{*}{First-/Third-party data}
& Purpose statement
& \cmark & \cmark & \cmark & \cmark & \cmark & \cmark \\
\cline{2-8}
& \revised{User choice}
& \xmark & \xmark & \cmark & \cmark & \xmark & \xmark \\
\hline

\multicolumn{2}{|l|}{\revised{Data sharing with other users}}
& \cmark & \cmark & \xmark & \cmark & \xmark & \cmark \\
\hline

\end{tabular}
\end{table*}




\begin{table*}[ht]
\renewcommand{\arraystretch}{1.2}
\small
\centering
\caption{Controllable units of different types of user data.}
\label{tab:DataUnits}
\begin{tabular}{|p{0.1\textwidth}|p{0.15\textwidth}|p{0.48\textwidth}|p{0.15\textwidth}|}
\hline
\multicolumn{2}{|c|}{\textbf{Data}} &
\multirow{2}{*}{\textbf{Description}} &
\multirow{2}{*}{\textbf{Applicable Platforms}} \\
\cline{1-2}
\textbf{Data Type} & \textbf{Data Units} & & \\
\hline

\multirow{9}{*}{Chat History}
& Model-generated image
& The image generated by Claude\revised{, which is} included in a message
& Claude \\
\cline{2-4}
& Individual message
& A piece of message sent by the user or generated by the model
& Character.ai, Claude, Gemini, Pi \\
\cline{2-4}
& Multiple messages
& Two or more individual messages from the user, the model, or both of them
& Pi \\
\cline{2-4}
& Single conversational round
& One message from the user followed by one reply generated by the model
& Gemini, Meta AI \\
\cline{2-4}
& Multiple conversational rounds
& Two or more back-and-forth turns of messages between the user and the model
& Gemini \\
\cline{2-4}
& Chat session
& A continuous message exchange between the user and the model, usually within a specific time frame or topic
& All platforms \\
\cline{2-4}
& Multiple chat sessions
& Two or more chat sessions between the user and the model
& Claude \\
\cline{2-4}
& Messages with a ``character''
& All conversation messages that the user has had with a specific ``character'' 
& Character.ai \\
\cline{2-4}
& All the chat history
& All conversation messages between the user and all models saved on the platform
& ChatGPT, Claude, Gemini, Meta AI \\
\hline

\multirow{2}{*}{Memory}
& Memory snippet
& A piece of information that the model derives from individual messages for personalizing \revised{interactions}, usually appears as a rephrased form of the original input. 
& ChatGPT, Gemini \\
\cline{2-4}
& ``Pinned'' Message
& A piece of message that users can ``pin'' to make the character \revised{remember} and \revised{personalize} future interactions with that character
& Character.ai \\
\hline

\multirow{3}{*}{\parbox[t]{\linewidth}{Customized\\Object}}
& Conversational agent (CA)
& The CA is customized by users with specified personas and communication styles---the \revised{``character''} in Character.ai, the \revised{``Gem''} in Gemini
& Character.ai, ChatGPT, Gemini \\
\cline{2-4}
& User persona
& A virtual role that the user wants to play during the interaction
& Character.ai \\
\cline{2-4}
& Project
& A specialized workspace for users to organize chats, upload documents, and create custom instructions to help streamline repetitive tasks and facilitate team collaboration
& Claude \\
\hline

\end{tabular}
\end{table*}


\subsection{Data Governance}~\label{Control_Policy}
As described in Section~\ref{method: governance}, we \revised{applied} a revised version of Wilson et al.'s privacy policy annotation scheme~\cite{Wilson2016Scheme} to analyze each platform's data governance practices based on their Privacy Policies and Terms of Service. Table~\ref{tab:governance} presents an overview of data governance on these aspects across the six platforms. 
Below, we \revised{describe} the governance \revised{practice that stood out among} conversational LLM platforms compared to traditional platforms, highlighting whether they grant users ownership of \revised{models'} generated data, what controls are available, and how shared data is governed among multiple users.

\subsubsection{Data Ownership} Except for Pi, which only mentions that users own their inputs, all other platforms explicitly grant users full ownership of both submitted content (e.g., input text, uploaded files, pictures, audio recordings) and platform-generated outputs, as stated in their Terms of Service or Terms of Use. Additionally, all platforms claim to offer users the right to access, update, correct, or delete their personal data, ensuring they have control over their information.

\subsubsection{First/Third-Party Data Usage} All platforms provide detailed information on data collection, usage, and users' rights to data control by describing the types of data collected and the purposes of data processing by first (i.e., the platform itself) and third parties (e.g., analytics tools, external reviewers, and vendors). The collected data includes the information directly provided by users (e.g., the email \revised{address} entered in the user's profile) as well as data automatically gathered through system interactions (e.g., the chat history). The primary purposes of both first and third-party data collection include service improvement, research and analysis, legal compliance, and personalized advertising.

\subsubsection{\revised{User Choice}}~\label{finding: opt-in-out} Except for Meta AI, which does not explicitly mention its practices about opt-out \revised{choice, other platforms claim that} users have the right to opt out of data usage involving both first and third parties on these platforms. 
Character.ai and Claude allow opt-out of targeted advertising through the third-party analytics tools they employ. Similarly, Pi \revised{claims that they offer} the opt-out option for online tracking by disabling third-party cookies.
ChatGPT and Gemini claim they offer opt-out options that allow users to prevent their data from being used for model training. 

\subsubsection{Data Sharing Among Multiple Users}~\label{findings: share-policy} Regarding data-sharing governance, Character.ai stands out as a community-driven platform, explicitly stating that ``\textit{popular characters}'' created by users may be retained even after account deletion. They state in their Privacy Policy:

``\textit{If a Character you (the user) create and set to `\textit{Public}' reaches a certain threshold of popularity, we (Character.ai) reserve the right to preserve that Character’s characteristics and to keep that Character active on the Services, even if you otherwise delete your data and your account. We do this to avoid impacting the experience of other users, given that a highly popular Character by definition is having active conversations with many thousands of users. [...]}.''

Gemini emphasizes user control over shared content but warns that public information may become searchable. ChatGPT allows users to share conversations and interact with third-party services, with shared data subject to external privacy policies. Other platforms did not explicitly mention their practices in the aforementioned documents.


         

\begin{table*}[t]
\centering
\small
\caption{\revised{The control options offered by the six conversational LLM platforms.~\cmark~refers to that the platform supports the specific control operation, while~\xmark~refers to that the platform does not support the specific control operation; \textbf{N/A} means that no other control options are available or the user data is not integrated in the corresponding form.}}
\label{tab:Overview}
\renewcommand{\arraystretch}{1.2}

\begin{tabular}{|
    p{0.13\textwidth}| 
    p{0.13\textwidth}| 
    p{0.16\textwidth}| 
    p{0.13\textwidth}| 
    p{0.07\textwidth}| 
    p{0.07\textwidth}| 
    p{0.07\textwidth}| 
    p{0.05\textwidth}|}
\hline
\multicolumn{2}{|c|}{\textbf{Control / Platform}} &
\multirow{2}{*}{\textbf{Character.ai}} &
\multirow{2}{*}{\textbf{ChatGPT}} &
\multirow{2}{*}{\textbf{Claude}} &
\multirow{2}{*}{\textbf{Gemini}} &
\multirow{2}{*}{\textbf{Meta AI}} &
\multirow{2}{*}{\textbf{Pi}} \\
\cline{1-2}
\textbf{Data type} & \textbf{Control option} & & & & & & \\
\hline

\multirow{6}{*}{Chat History}
  & Access   & \cmark & \cmark & \cmark & \cmark & \cmark & \cmark \\ \cline{2-8}
  & Retrieve & \cmark & \cmark & \cmark & \cmark & \cmark & \cmark \\ \cline{2-8}
  & Edit     & \cmark & \xmark & \cmark & \cmark & \cmark & \xmark \\ \cline{2-8}
  & Delete   & \cmark & \cmark & \cmark & \cmark & \cmark & \xmark \\ \cline{2-8}
  & Share    & \xmark & \cmark & \cmark & \cmark & \cmark & \cmark \\ \cline{2-8}
  & Others   & ``Remove'', export, and archive &
               Export and (un)archive &
               Export & N/A & Export & Export \\
\hline

\multirow{5}{*}{Memory}
  & ``Memorize'' & \cmark~GUI & \cmark~NL &
    \multirow{5}{*}{N/A} &
    \cmark~NL, GUI &
    \multirow{5}{*}{N/A} &
    \multirow{5}{*}{N/A} \\ \cline{2-4}\cline{6-6}
  & Access   & \cmark~GUI & \cmark~GUI &  & \cmark~GUI &  &  \\ \cline{2-4}\cline{6-6}
  & Retrieve & \cmark~NL  & \cmark~NL  &  & \cmark~NL  &  &  \\ \cline{2-4}\cline{6-6}
  & ``Update memory'' & \cmark~GUI & \cmark~NL &  &
    \cmark~NL, GUI &  &  \\ \cline{2-4}\cline{6-6}
  & ``Forget'' (or delete) & \cmark~GUI &
    \cmark~NL, GUI &  &
    \cmark~NL, GUI &  &  \\
\hline

\multirow{6}{*}{Customized Object}
  & Access   & \cmark & \cmark & \cmark & \cmark &
    \multirow{6}{*}{N/A} &
    \multirow{6}{*}{N/A} \\ \cline{2-6}
  & Retrieve & \cmark & \cmark & \cmark & \cmark & & \\ \cline{2-6}
  & Edit     & \cmark & \cmark & \cmark & \cmark & & \\ \cline{2-6}
  & Delete   & \cmark & \cmark & \cmark & \cmark & & \\ \cline{2-6}
  & Share    & \cmark & \cmark & \xmark & \cmark & & \\ \cline{2-6}
  & Others   & ``Remove'' & N/A & Archive & N/A & & \\
\hline

\end{tabular}
\end{table*}


\vspace{2mm}
\subsection{Controllable Data and Options}~\label{WhatControl}

Through our walkthrough, we observed three types of data users can control: \revised{Chat} history, memory, and customization objects. However, the data units are structured differently across platforms, as summarized in Table~\ref{tab:DataUnits}. 
As mentioned in Section~\ref{terms}, we define data units as discrete elements of data for control, which are building blocks for platforms to implement controls on users' data. 
The options that users can perform control on these data (e.g., access, edit, delete, share) also vary by platforms (see Table~\ref{tab:Overview}). 
Below, we describe these units in detail alongside the control options that can be performed over them. For ease of reading, data units appear in \dataunit{bold small caps}, control options are \controloptions{underlined}, and \textit{italics} text refers to words or phrasing directly quoted from the platforms' interfaces or their institutional materials.


\begin{figure*}[t]
    \centering
    \includegraphics[width=\linewidth]{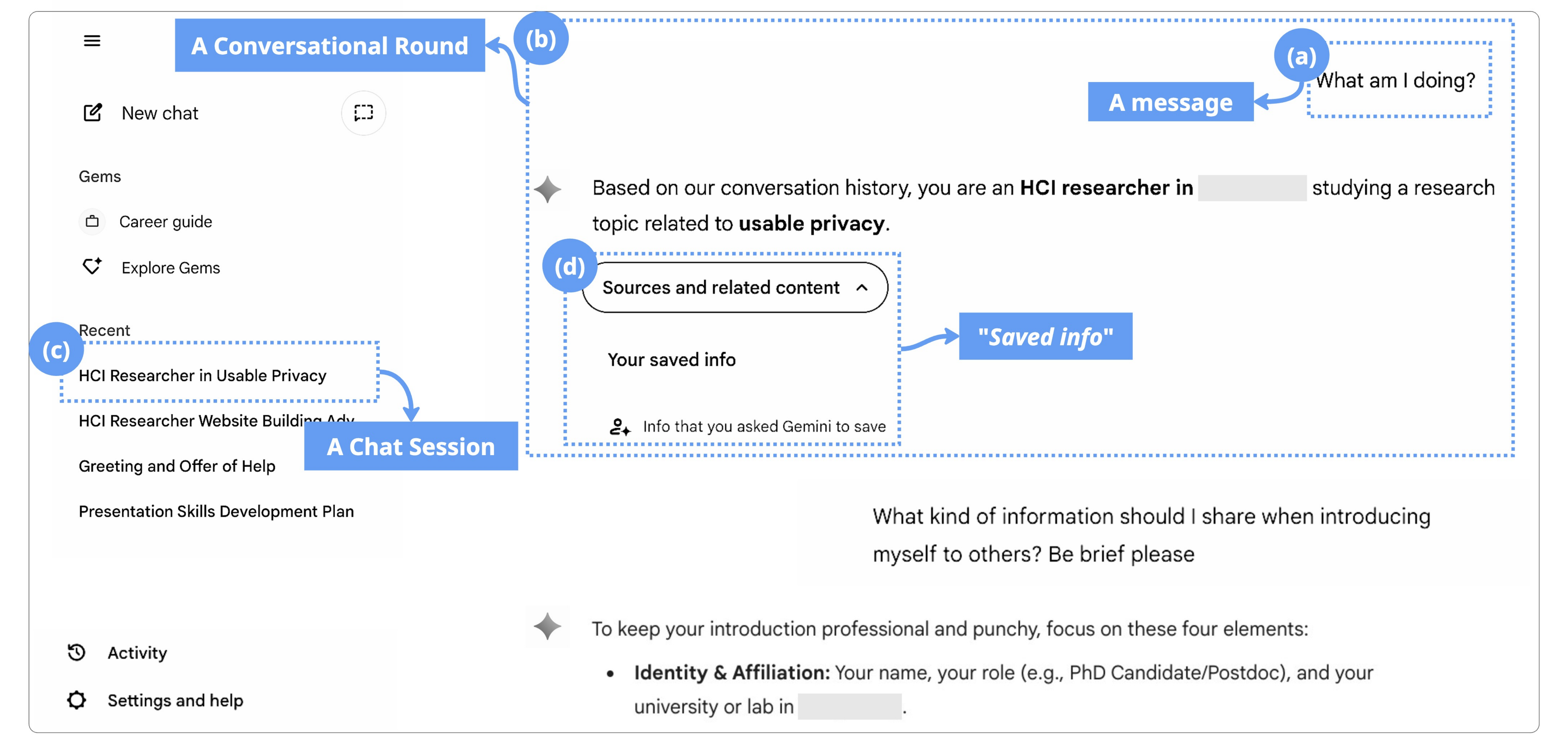}
    \Description{In Gemini, the data units of chat history include (a) a message, (b) a conversational round, and (c) a chat session. (d) When a message generated by Gemini is based on ``\textit{saved info}'' (i.e., memory), a notification appears indicating which memory snippet was referenced.}
    \caption{\revised{In Gemini, the data units of chat history include~{\raisebox{-0.5ex}{\includegraphics[height=1.2em]{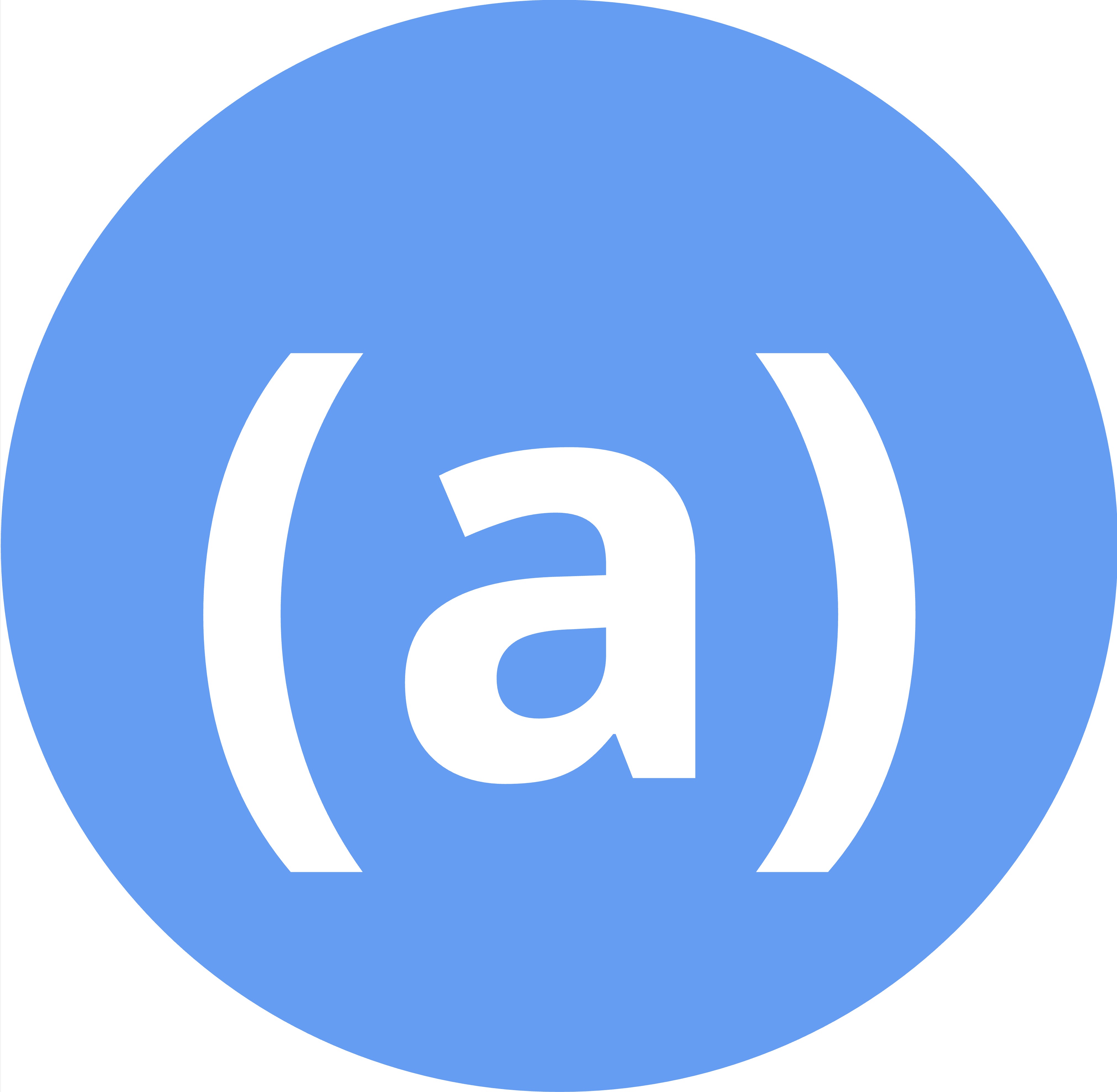}}}~a message,~{\raisebox{-0.5ex}{\includegraphics[height=1.2em]{Figures/a.jpg}}}~a conversational round, and~{\raisebox{-0.5ex}{\includegraphics[height=1.2em]{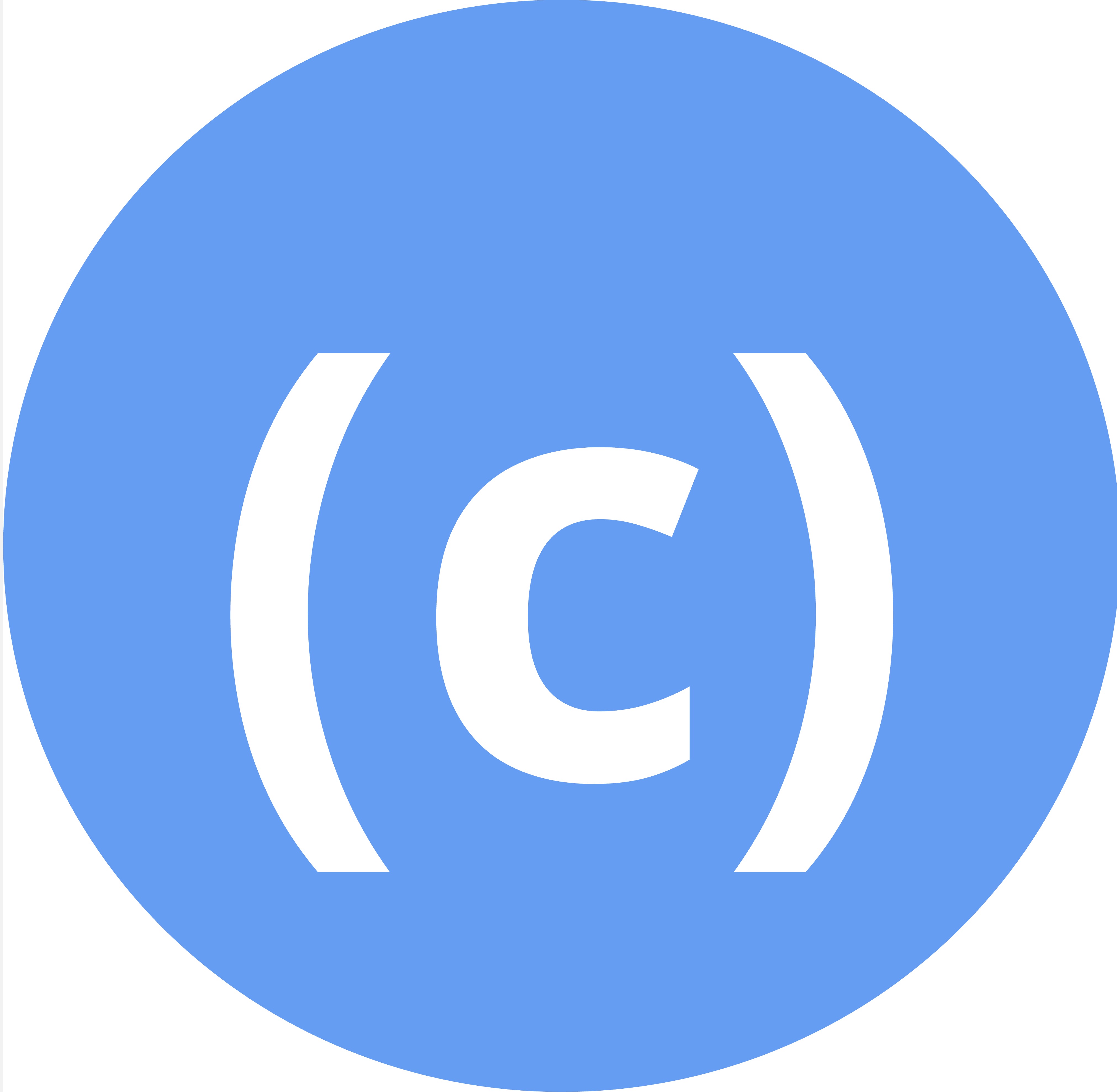}}}~a chat session.~{\raisebox{-0.5ex}{\includegraphics[height=1.2em]{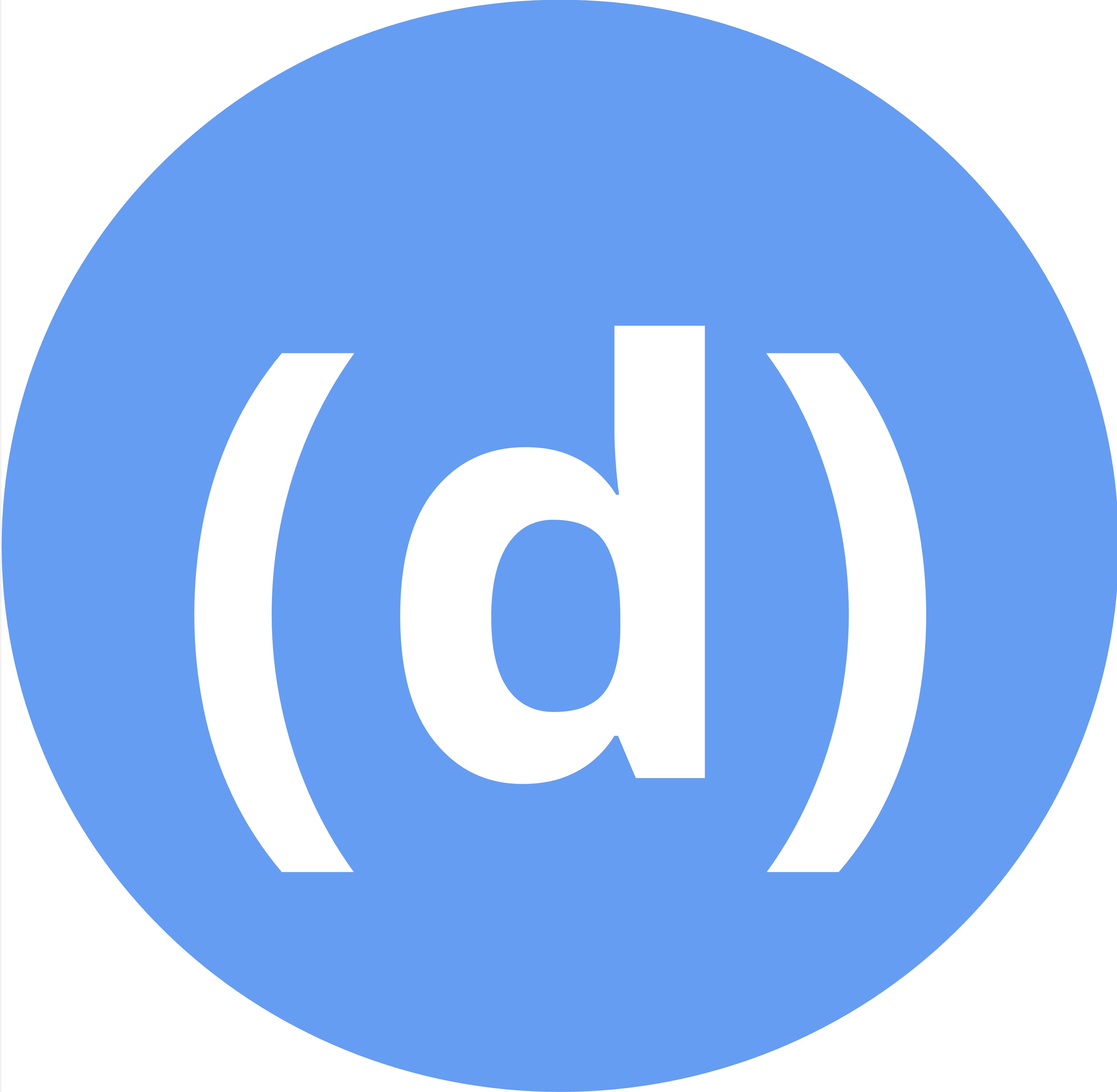}}}~When a message generated by Gemini is based on ``\textit{saved info}'' (i.e., memory), a notification appears indicating which memory snippet was referenced.}}
    \label{fig: ScreenChatDataUnit}
\end{figure*}


\subsubsection{\revised{Chat History: Varying Data Units From Message to Session}}~\label{Chat History}
As the central component of user data, chat history refers to recorded conversational interactions between users and LLMs, which consist of users' inputs and the LLMs' responses. We found that each of the six platforms aggregates chat history differently, varying from message, conversational round to chat session, see all data units and their descriptions in \revised{Table~\ref{tab:Overview}}.

The most basic unit is~\dataunit{a message}, either a single user input or a single model output. Building on this,~\dataunit{a conversational round} \revised{contains} a user input with \revised{corresponding model} response, while~\dataunit{a} \dataunit{chat session} (``\textit{Threads}'' in Meta AI and Pi) encompasses a continuous sequence of exchanges within a particular context or time frame,~\revised{see Figure~\ref{fig: ScreenChatDataUnit}~{\raisebox{-0.5ex}{\includegraphics[height=1.2em]{Figures/a.jpg}}}$\rightarrow$~{\raisebox{-0.5ex}{\includegraphics[height=1.2em]{Figures/c.jpg}}}.} Sessions are typically initiated when users select ``\textit{Start a new chat session}.''  At the broadest level, some platforms allow users to control~\dataunit{all chat history} collectively, treating them as a bulk dataset. 

\begin{figure*}
    \centering
    \includegraphics[width=\linewidth]{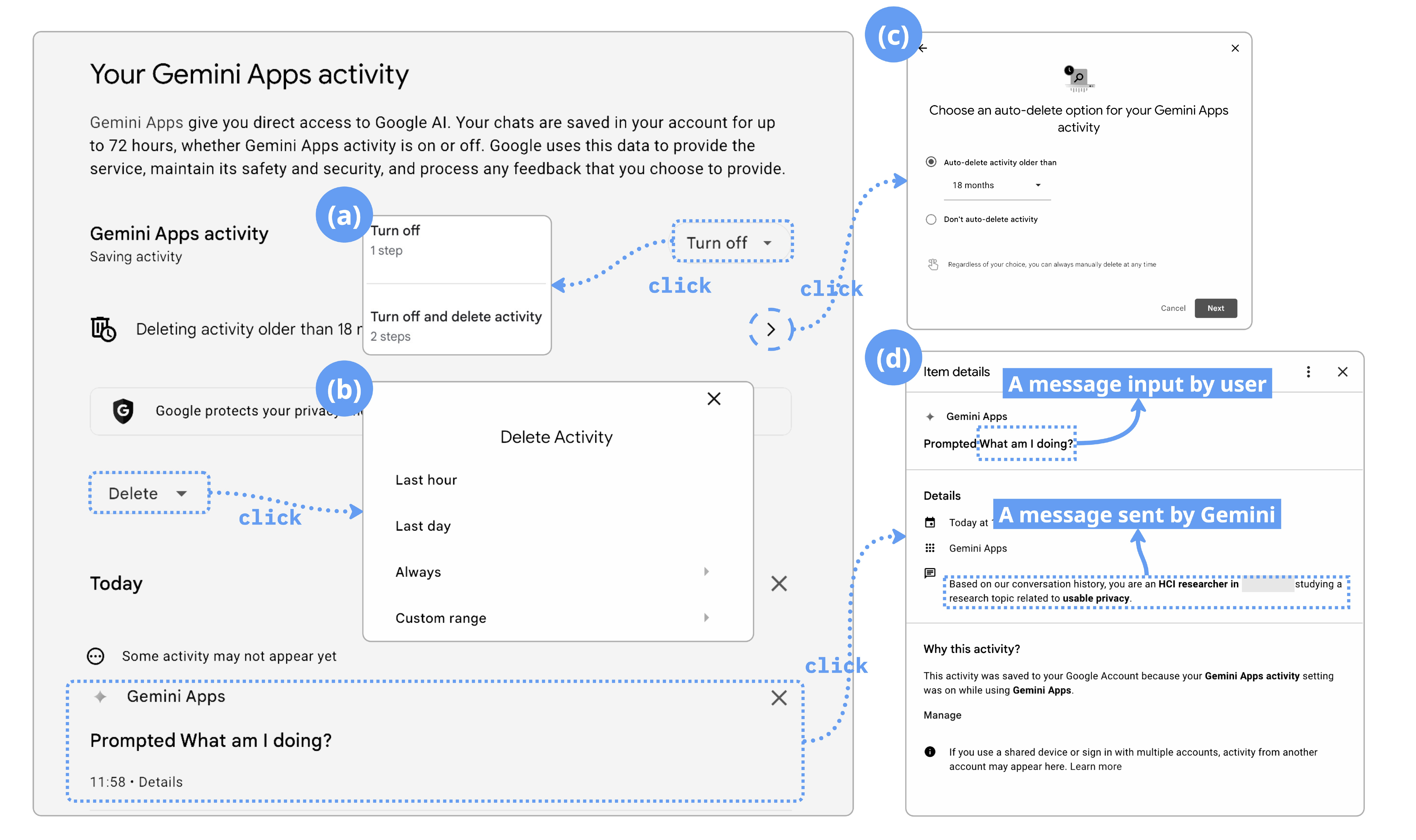}
    \Description{The “Gemini Apps Activity” page lets users manage their interaction history with Gemini. Users can choose “Turn off” or “Turn off and delete activity.” Turning off activity prevents future chats from appearing in the activity log and from being used to train the models. Choosing “Turn off and delete activity” also deletes all chat history. Users can manually delete individual conversational rounds or multiple rounds within ``Last hour,'' ``Last day,'' ``Always,'' or a ``Custom range.'' Users can enable auto-deletion within the Gemini app, with options for “3 months,” “18 months” (default), “36 months,” or “Not auto-delete activity.” Chat history is grouped under “Item Details,” which represents a conversational round containing one user input and one Gemini output.}
    \caption{\revised{The ``\textit{Gemini Apps Activity}'' page, where users can manage their interaction history with Gemini.~{\raisebox{-0.5ex}{\includegraphics[height=1.2em]{Figures/a.jpg}}}~Users can choose to ``\textit{Turn off}'' or ``\textit{Turn off and delete activity}''. Turning off activity prevents future chats from appearing in the activity log and from being used to train the models. Choosing ``\textit{Turn off and delete activity}'' also deletes all chat history.~{\raisebox{-0.5ex}{\includegraphics[height=1.2em]{Figures/a.jpg}}}~Users can manually delete individual conversational rounds or multiple rounds within ``\textit{Last hour},'' ``\textit{Last day},'' ``\textit{Always},'' or a ``\textit{Custom range}.''~{\raisebox{-0.5ex}{\includegraphics[height=1.2em]{Figures/c.jpg}}}~Users can enable auto-deletion within the Gemini app, with options for ``\textit{3 months},'' ``\textit{18 months}'' (default), ``\textit{36 months},'' or ``\textit{Not auto-delete activity}.''~{\raisebox{-0.5ex}{\includegraphics[height=1.2em]{Figures/d.jpg}}}~Chat history is grouped under ``\textit{Item Details},'' which represents a conversational round containing one user input and one Gemini output.}}
    \label{fig: ScreenGeminiActivity}
\end{figure*}

As summarized in Table~\ref{tab:Overview}, all platforms allow users to \controloptions{access} and \controloptions{retrieve} their chat history. Except for Gemini, they also support \controloptions{exporting} all of the user's chat history. However, the availability of other control options, as well as the granularity of the data units over which these controls can be exercised, varies across platforms:
\begin{itemize}
\item \controloptions{Editing}: All platforms except ChatGPT and Pi allow users to edit the most recent user message. Character.ai additionally permits users to edit the last response from the model. Yet no platform currently supports editing earlier messages or full chat sessions.

\item \controloptions{Deletion}: With the exception of Character.ai, all platforms do not allow users to delete individual messages. ChatGPT, Claude, Gemini, and Meta AI support \revised{the} deletion of chat histories by chat sessions. Claude further \revised{provides the deletion option} of multiple sessions via keyword search. Gemini stands out by offering the option to delete~\dataunit{single} or~\dataunit{multiple conversational rounds} within a selected time frame through its ``\textit{Gemini Apps Activity}'' interface,~\revised{see Figure~\ref{fig: ScreenGeminiActivity}~{\raisebox{-0.5ex}{\includegraphics[height=1.2em]{Figures/d.jpg}}}~and~{\raisebox{-0.5ex}{\includegraphics[height=1.2em]{Figures/a.jpg}}}}. At a global level, ChatGPT, Claude, Gemini, and Meta AI allow users to delete all chat history from account settings. 

\item \controloptions{Sharing}:
Character.ai does not allow users to share chat history at any level. ChatGPT and Gemini support sharing entire chat sessions. Pi provides more flexible sharing, allowing users to select and share multiple non-sequential messages (including either user inputs, model outputs, or both). Meta AI allows sharing by conversational round, meaning each shared unit must include a user input and the corresponding model response.
\revised{In addition,} Claude supports the sharing of generated artifacts (e.g., images created by the model) independently of the chat context in which they were produced. These are treated as standalone outputs, suggesting Claude differentiates between conversational content and generative media.

 \end{itemize}
 
\subsubsection{Memory: A Type of Derived User Data}~\label{Units_Memory}


Memory is a unique type of data on conversational LLM platforms, available on ChatGPT and Gemini. Unlike the original user input, memories are ``derived'' data that the model considers meaningful for personalizing user interaction. As ChatGPT's ``\textit{Memory FAQ}'' states, ``\textit{Memory works similarly to Custom instructions, [...] when you share information that might be useful for future conversations, we’ve trained the model to add a summary of that information to its memory.}'' In this sense, memory represents a more complex integration of information accumulated over time, rather than a simple collection of individual messages or chat sessions.

Our analysis found that there are no distinct, pre-defined units of memory. Instead, both ChatGPT and Gemini store memory as pieces of information reflecting the models’ evolving understanding of a user, which we term ``\dataunit{memory snippet}.'' In short, a memory snippet is a specific data unit that captures what the model decides to keep and use from a user-input message, usually in a shortened or rephrased form. \revised{We also} found that each memory snippet is derived from a single message but may contain one or multiple distinct facts or attributes, such as ``\textit{I like blue}'' and ``\textit{My name is ...}'', depending on what information is considered important to remember by the model.
Memory snippets function as contextual information across chat sessions. In other words, snippets generated in one session can be retrieved, edited, and deleted in other sessions.

\begin{figure*}[t]
    \centering
    \includegraphics[width=\linewidth]{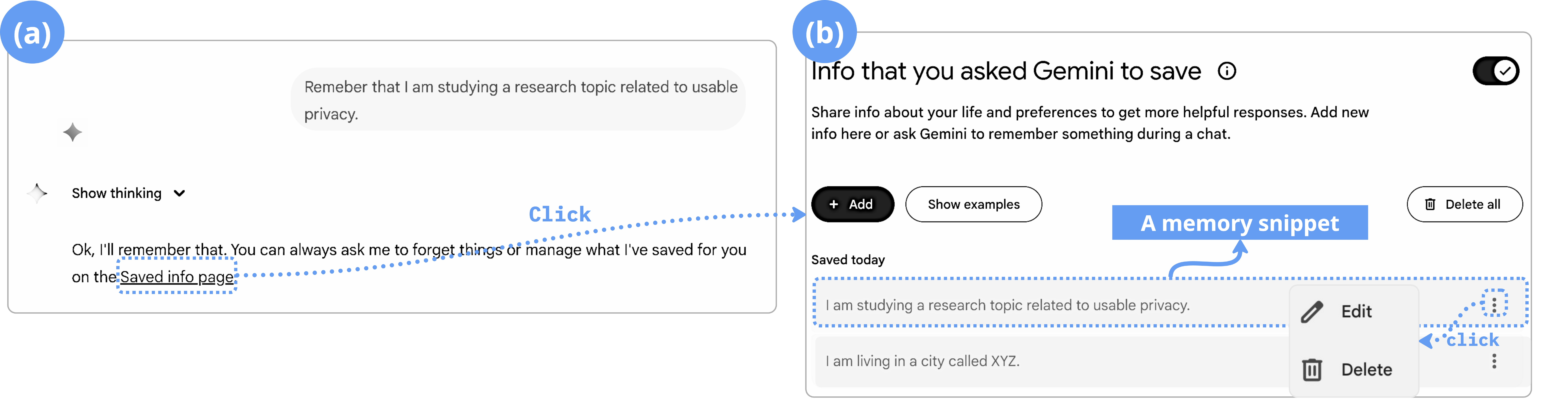}
    \Description{Gemini’s memory snippets and Saved Info portal. Using a prompt such as “Remember ...”, users can instruct Gemini to save information as memory. Saved information is displayed in the “Info that you asked Gemini to save” page, where the memory is organized by snippets (pieces of information). Users can access, edit, or delete individual snippets, and can also delete all snippets from this interface.}
    \caption{\revised{Gemini's memory snippets and Saved Info portal.~{\raisebox{-0.5ex}{\includegraphics[height=1.2em]{Figures/a.jpg}}}~Using a prompt such as ``\textit{Remember ...},'' users can instruct Gemini to save information as memory.~{\raisebox{-0.5ex}{\includegraphics[height=1.2em]{Figures/a.jpg}}}~Saved information is displayed in the ``\textit{Info that you asked Gemini to save}'' page, where the memory is organized by snippets (pieces of information). Users can access, edit, or delete individual snippets, and can also delete all snippets from this interface.}}
    \label{fig: ScreenGeminiMemory}
\end{figure*}

\begin{figure*}
    \centering
    \includegraphics[width=\linewidth]{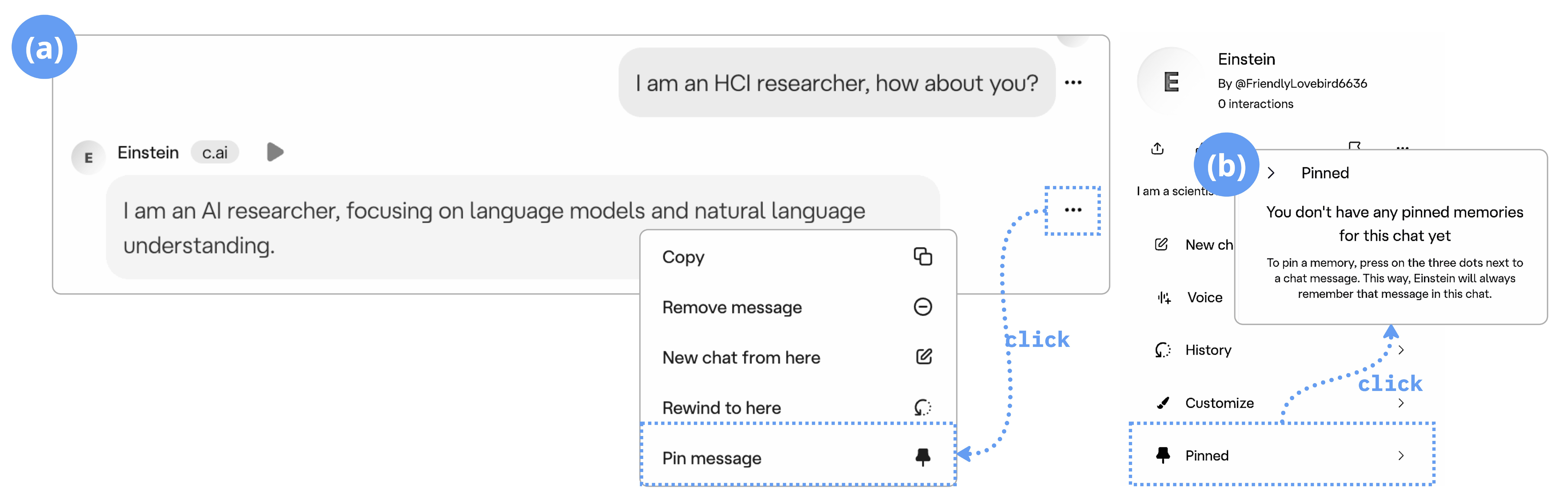}
    \Description{On Character.ai, users can “pin” messages, and the character will always remember that message in the chat. These pinned messages appear in the right-side panel next to the conversational interface.}
    \caption{\revised{On Character.ai,~{\raisebox{-0.5ex}{\includegraphics[height=1.2em]{Figures/a.jpg}}}~ users can ``\textit{pin}'' messages, then the character ``\textit{will always remember that message in this chat}.''~{\raisebox{-0.5ex}{\includegraphics[height=1.2em]{Figures/a.jpg}}}~These pinned messages will be displayed in the right side panel beside the CUI.}}
    \label{fig: ScreenPin}
\end{figure*}

As shown in Figure~\ref{fig: ScreenGeminiMemory}~{\raisebox{-0.5ex}{\includegraphics[height=1.2em]{Figures/a.jpg}}}~and~{\raisebox{-0.5ex}{\includegraphics[height=1.2em]{Figures/a.jpg}}}, memory is created and controlled in the form of memory snippets on ChatGPT and Gemini, where users can \controloptions{access}, \controloptions{edit}, \controloptions{delete} these memory snippets.
On Character.ai, while the term memory is not used, we observed a feature similar to memory: users can ``\controloptions{\textit{pin}}'' specific messages and ask the character to ``memorize'' them, see \revised{Figure~\ref{fig: ScreenPin}~{\raisebox{-0.5ex}{\includegraphics[height=1.2em]{Figures/a.jpg}}}}. These ``\dataunit{pinned \revised{messages}}'' will then be the controllable data unit of memory on Character.ai,~\revised{see Figure~\ref{fig: ScreenPin}~{\raisebox{-0.5ex}{\includegraphics[height=1.2em]{Figures/a.jpg}}}}. Vice versa, they can also ``\controloptions{\textit{unpin}}'' these messages so that the CA will ``\textit{forget}'' the information and update its memory accordingly.

Beyond simply reviewing the saved memory snippets, Character.ai, ChatGPT, and Gemini allow users to \controloptions{retrieve} information from memory, which refers to the model actively bringing up previously ``remembered'' information to an ongoing conversation. 
For example, if a user sends a message, ``\textit{My name is Johnny},''  and later inquires ``\textit{What is my name?},'' the model can then retrieve from its memory and send a response like ``\textit{Your name is Johnny}.'' 
Note that such retrieval is not limited to direct user inquiries but can also occur implicitly to support personalization. In another example, if a user once shared, ``\textit{I prefer concise answers},'' the model may adjust its response style in future interactions without needing an explicit prompt.

\subsubsection{Customized object:~\revised{An} Encapsulated Data Unit}~\label{customization}


On these conversational LLM platforms, ``customization'' refers to user-initiated adjustments in how the model responds to their input, which differs from the customization of GUI components such as color mode or font size on traditional platforms. \revised{We use the term customized object for a set of platform-specific features that share the same structure:} they \revised{encapsulate user-provided} textual instructions, documents, images, or audio, to form a single data unit, with unique instructions to shape the model responses. \revised{Across platforms, these customized objects manifest in three primary forms: conversational agents (CA), user personas, and projects.}


\begin{figure*}
    \centering
    \includegraphics[width=\linewidth]{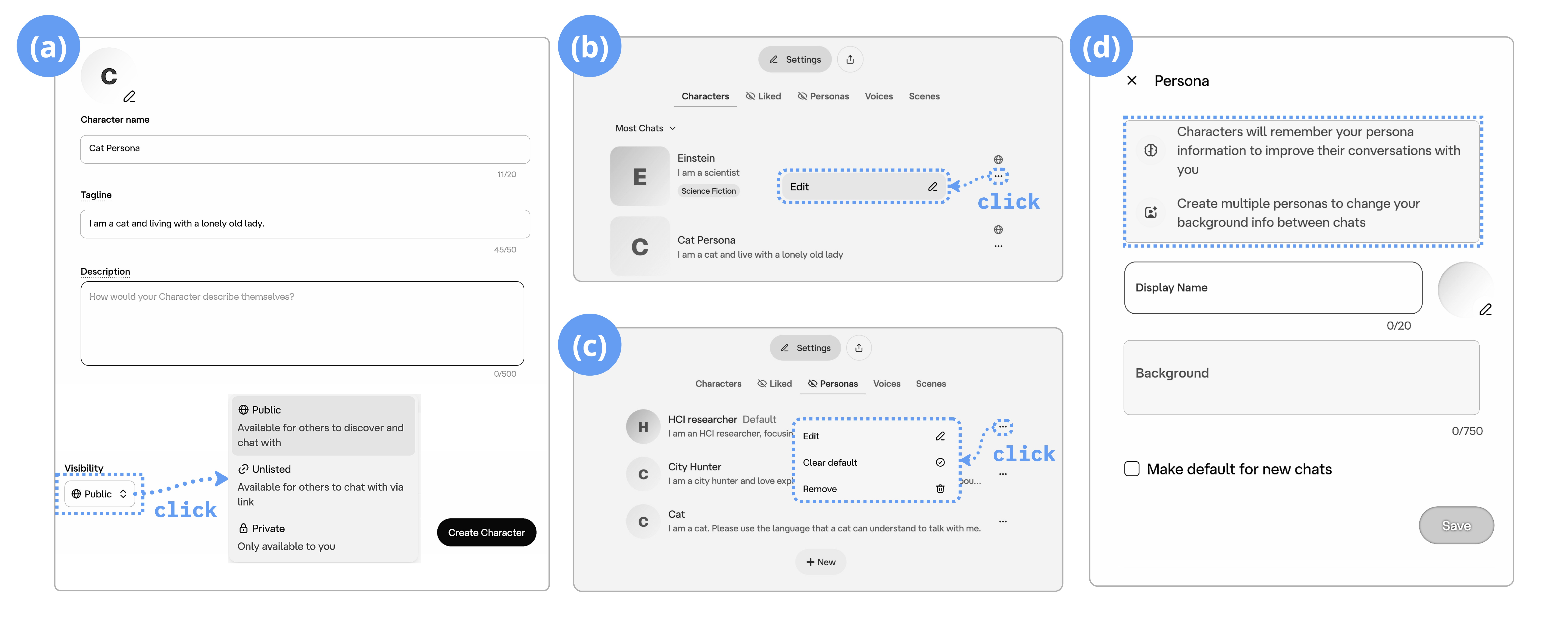}
    \Description{Create, access, edit, delete, and share customized objects on Character.ai. Users can create a customized character by entering a description and choosing its visibility settings: “private,” “unlisted,” or “public for others to discover and chat with.” Customized characters can be edited through the Characters section in Settings. Through the Settings page under the Personas section, users can edit a persona, set or clear a default persona for new chats, or remove a customized persona. Users can create multiple personas to change their background info between chats, and characters will remember the user’s persona information to improve conversations.}
    \caption{\revised{Creating, accessing, editing, deleting, and sharing customized objects on Character.ai.~{\raisebox{-0.5ex}{\includegraphics[height=1.2em]{Figures/a.jpg}}}~Users can create a customized character by entering a description and choosing its visibility settings~``\textit{private},'' ``\textit{unlisted},'' or ``\textit{public for others to discover and chat with}.''~{\raisebox{-0.5ex}{\includegraphics[height=1.2em]{Figures/a.jpg}}}~Customized characters can be edited through the ``\textit{Characters section}'' in Settings.~{\raisebox{-0.5ex}{\includegraphics[height=1.2em]{Figures/c.jpg}}}~Through the Settings page under the section ``\textit{Personas},'' users can edit, set or clear a persona as default for new chats, or remove the customized persona.~{\raisebox{-0.5ex}{\includegraphics[height=1.2em]{Figures/d.jpg}}}~Users can ``\textit{create multiple personas to change their background info between chats},'' and ``\textit{characters will remember the user’s persona information to improve the conversations}.''}}
    \label{fig: ScreenCustomization}
\end{figure*}


\revised{Among the six platforms, Character.ai, ChatGPT, Claude, and Gemini support users to customize model responses, often in the form of specifying the identity and communications styles of the~\dataunit{customized CA}, including the \textit{character} in Character.ai (see Figure~\ref{fig: ScreenCustomization}~{\raisebox{-0.5ex}{\includegraphics[height=1.2em]{Figures/a.jpg}}}~and~{\raisebox{-0.5ex}{\includegraphics[height=1.2em]{Figures/a.jpg}}}), \textit{GPT} in ChatGPT, \textit{Gem} in Gemini. Character.ai also \revised{offers} customization for~\dataunit{user personas}, a virtual role that users want to play during their interactions with the character,~\revised{see Figure~\ref{fig: ScreenCustomization}~{\raisebox{-0.5ex}{\includegraphics[height=1.2em]{Figures/a.jpg}}}~and~{\raisebox{-0.5ex}{\includegraphics[height=1.2em]{Figures/d.jpg}}}}.}

For example, on ChatGPT, a customized GPT may include a role-defining instruction (e.g., ``\revised{\textit{You are my personal writing assistant.}}''), a supplementary file (e.g., a writing sample), a profile image, and a selected voice. It is noteworthy that, for each customized object, we see its properties such as avatar, voice, textual descriptions, and uploaded files as part of the object, rather than controllable units. 

\begin{figure*}
    \centering
    \includegraphics[width=\linewidth]{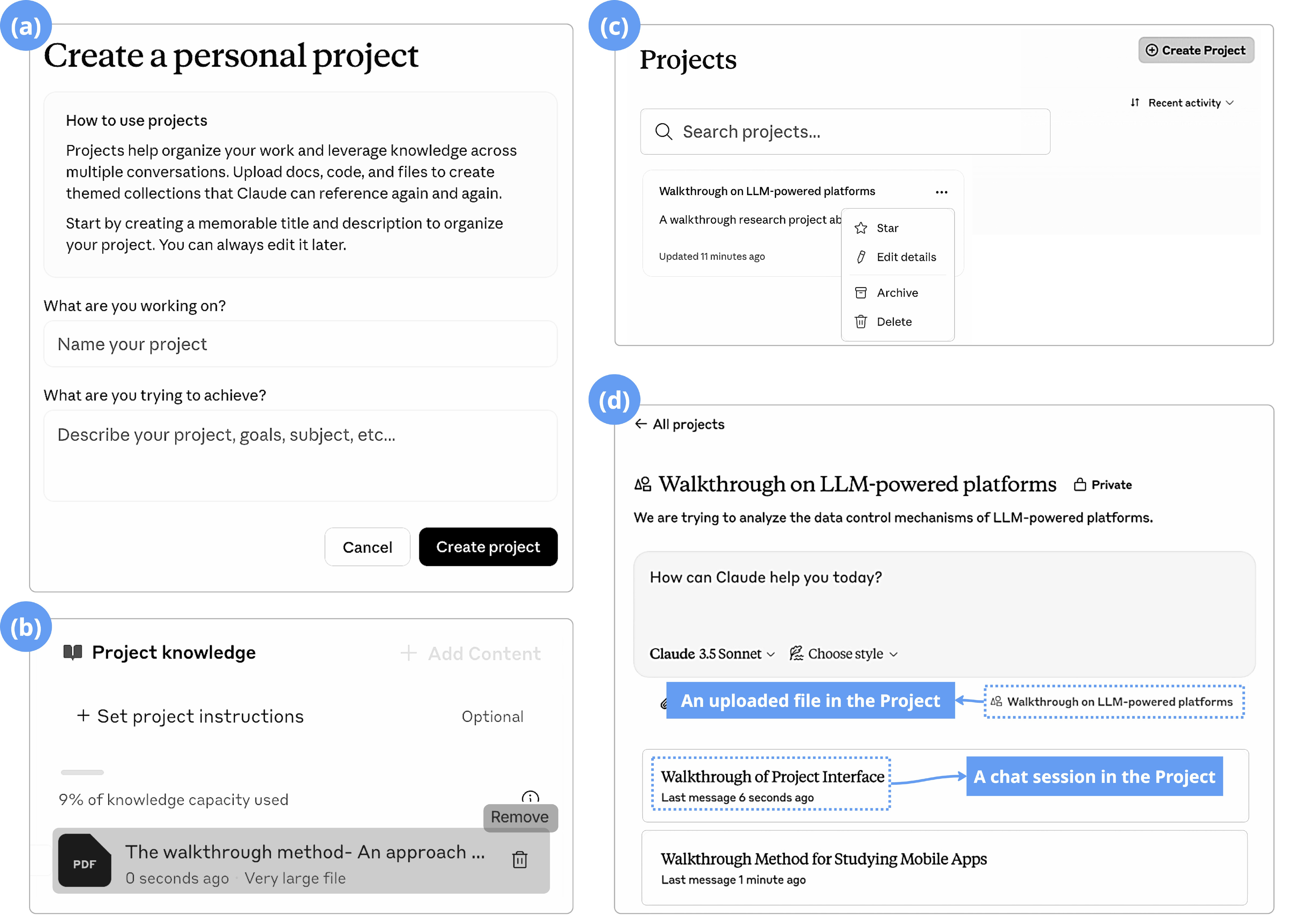}
    \Description{Claude allows users to create, access, edit, delete, and share projects. Users can create a project by providing a project title and description. They can customize projects by uploading documents, code, and other files to a project’s knowledge base to create themed collections that Claude can reference repeatedly. Users can access customized projects through the Projects Portal, and they can initiate multiple chat sessions within a project.}
    \caption{\revised{Claude allows users to create, access, edit, delete, and share projects.~{\raisebox{-0.5ex}{\includegraphics[height=1.2em]{Figures/a.jpg}}}~Users can create a project by providing project title and description.~{\raisebox{-0.5ex}{\includegraphics[height=1.2em]{Figures/a.jpg}}}~Users can customize projects by ``\textit{uploading documents, code, and other files} to a project's knowledge base for ``\textit{creating themed collections that Claude can reference again and again}.''~{\raisebox{-0.5ex}{\includegraphics[height=1.2em]{Figures/c.jpg}}}~Users can access customized projects through the ``Projects Portal.''~{\raisebox{-0.5ex}{\includegraphics[height=1.2em]{Figures/d.jpg}}}~Users can initiate multiple chat sessions within a project.}}
    \label{fig: ScreenProject}
\end{figure*}

Besides customized CAs and user personas, users can customize \dataunit{projects} on Claude; projects are specialized workspaces for users to organize chats, upload documents, and create custom instructions, which helps streamline repetitive tasks and facilitate team collaboration, see Figure~\ref{fig: ScreenProject}~{\raisebox{-0.5ex}{\includegraphics[height=1.2em]{Figures/a.jpg}}} $\rightarrow${\raisebox{-0.5ex}{\includegraphics[height=1.2em]{Figures/d.jpg}}}.
All of these customized objects can contain rich information about users. For instance, customized CAs and user personas may reveal sensitive information about users' identities and behaviors, while projects can include confidential documents and instructions.


Regarding controls, the four platforms that \revised{allow} customization of model responses all support users in \controloptions{accessing} and \controloptions{editing} the content in their customized objects (avatar, voice of customized characters, files in customized projects), with other varying control options described below.
\begin{itemize}
    \item \controloptions{Deletion}: 
    ChatGPT, Claude, and Gemini allow users to delete the customized object (e.g., CA persona, projects, and response style). 
    Character.ai, however, only allows deletion of customized user personas but not customized ``\textit{Characters},'' \revised{see Figure~\ref{fig: ScreenCustomization}~{\raisebox{-0.5ex}{\includegraphics[height=1.2em]{Figures/a.jpg}}}~and~{\raisebox{-0.5ex}{\includegraphics[height=1.2em]{Figures/c.jpg}}}.}
    Additionally, ChatGPT allows users to delete historical versions of customized CAs that are chronologically organized. 

    \item \controloptions{Sharing}: Character.ai, ChatGPT, and Gemini allow users to share customized CAs but not customized user personas; Claude does not provide a sharing option for customized projects on personal accounts. These platforms also offer different sharing mechanisms, which will be further detailed in Section~\ref{Co-ownData}. 

    \item \controloptions{Retrieval}: Similar to memory, the information about customized objects can be retrieved. For example, if a user customizes a CA as ``\textit{You are the philosopher Aristotle. You have wisdom on observing the world and providing provocative insights}\revised{,'' users} can retrieve the information by inquiring ``\textit{Who are you?}'' and ``\textit{How can you help me?}'' during the conversations with the customized character.
\end{itemize}

\subsection{Control Execution}
Building on the above section on what can be controlled and the associated control options, this section elaborates on how users could execute these control options (interaction method), when they can execute the control (before or after data input), the consequences of their control execution (scope of effects), and control executions with shared data (rights of the sharers and sharees).

\subsubsection{Graphic User Interface vs. Natural Language Control}~\label{Control_Memory}
Our walkthrough showed that all the control options mentioned above (except for retrieving memory or information about customized objects) can be executed at the graphic user interface (GUI) level. Similar to controls on traditional digital platforms~\cite{Bemmann2024Slider}, these controls are embedded as visible interface elements such as in-conversation menus, settings pages, and side panels, and are enacted through familiar mechanisms, such as direct text-field edits, toggles, checkboxes, drop-down menus, buttons, etc.
On ChatGPT, for example, users can access memory snippets through the memory portal and directly edit or delete these snippets (see Figure~\ref{app:interfaces}). 

It stands out that ChatGPT and Gemini also enable natural language (NL) control for memory and customized objects, most commonly, by instructing the model to ``\textit{remember}'' or ``\textit{forget}'' certain information in a chat session. 
On ChatGPT, there is explicit guidance on this approach under the``{Personalization}'' tab of the Settings page: ``\textit{To understand what ChatGPT remembers or teach it something new, just chat with it:
`Remember that I like concise responses.'
`I just got a puppy!' [...]}.''
Gemini provides similar instructions, with prompt examples to guide users on how to influence the CA's memory. As indicated by their instructions, the prompts are rather intuitive without fixed commands to follow. 
As mentioned in Section~\ref{Units_Memory} and~\ref{customization}, besides the explicit snippets in the memory portal, users can also retrieve the information from their previous interactions by NL. 

To inform users what is being remembered during the conversation, both ChatGPT and Genimi incorporate visual indicators. For instance, in ChatGPT, when actions such as ``\textit{memorizing,}'' ``\textit{editing memory,}'' or ``\textit{forgetting}'' are triggered, a small widget labeled ``\textit{memory updated}'' would appear at the top of the model response, which can be clicked and show the detailed memory snippet, see Figure~\ref{app:interfaces}. In Gemini, when the message sent by Gemini is referenced from memory, there is a label attached at the end of the message ``\textit{Sources and related content},'' where users are made aware of the referred memory, \revised{see Figure~\ref{fig: ScreenChatDataUnit}~{\raisebox{-0.5ex}{\includegraphics[height=1.2em]{Figures/d.jpg}}}}. 

\subsubsection{Retrospective vs. Proactive Control} 
All the platforms granted the right of retrospective control, such as access, edit, and deletion (see Table~\ref{tab:governance}), meaning users can act on their data~\textbf{after} it has already been entered or generated. When it comes to memory control, these retrospective options give users a chance to ``correct'' or delete the data that exists, but they cannot prevent it from being created in the first place. In other words, users do not know in advance whether or what memory snippets will be created.

In contrast, proactive control refers to mechanisms that allow users to set boundaries on data collection or usage~\textbf{before} their data is processed. 

On ChatGPT, there is a ``\textit{Temporary Chat}'' mode (see Figure~\ref{app:interfaces}), where user inputs are neither stored in memory nor used for model training, as stated in the conversation page:
``\textit{You’re in control of ChatGPT’s memory. You can reset it, clear specific or all memories, or turn this feature off entirely in your settings. If you’d like to have a conversation without memory, use Temporary Chat.}''
Gemini offers auto-deletion options, allowing users to configure the interaction history to be automatically deleted after a certain period of time, see Figure~\ref{fig: ScreenGeminiActivity}~{\raisebox{-0.5ex}{\includegraphics[height=1.2em]{Figures/c.jpg}}}. 
\revised{Gemini also provides a combined form of retrospective and proactive control through the ``\textit{Gemini Apps Activity}'' page, see Figure~\ref{fig: ScreenGeminiActivity}~{\raisebox{-0.5ex}{\includegraphics[height=1.2em]{Figures/a.jpg}}}. Users may choose ``\textit{Turn off}'' activity, which proactively prevents future chats from being logged or used for model training, or ``\textit{Turn off and delete activity},'' which additionally executes a retrospective deletion of all existing chat history.}

Additionally, as reported in Section~\ref{finding: opt-in-out}, on ChatGPT and Gemini, users can opt out of first-party uses for model training under their account settings, although it would not prevent the model from ``memorizing'' information deemed useful for personalizing further conversations.
However, the other four platforms do not provide similar control mechanisms for users to proactively manage their data, while they allow users to opt out of third-party data sharing, such as for advertising or analytics purposes. 

\subsubsection{Local vs. Global Effects}~\label{remove-semantics}
With the introduction of memory features, conversational LLM platforms no longer treat control operations as uniform. The seemingly same commands---such as delete, forget, or remove---can apply at different layers of the system, producing either \textit{\textbf{local effects}} (affect only the current chat session or information presented on the current interface) or \textit{\textbf{global effects}} (changes across all user data stored by the platform). 
In our walkthrough, only one type of control action consistently produces global effects, which is the ``delete all chat history'' option available in the account settings of ChatGPT, Claude, Gemini~\revised{(see Figure~\ref{fig: ScreenGeminiActivity}~{\raisebox{-0.5ex}{\includegraphics[height=1.2em]{Figures/a.jpg}}})}, and Meta AI. By contrast, most other controls produce only local effects.


In ChatGPT, for example, deleting a chat session has only a local effect: the conversation disappears from the side panel, but any information stored in memory remains. Similarly, when a user deletes a memory snippet through the memory portal, the stored information is removed from memory, while the original chat transcript remains accessible unless deleted separately. When a user attempts to delete a chat session, there will be a pop-up message: ``\textit{This will delete
<automatically generated chat session title>. To clear any memories from this chat, visit your settings.}'' 
Vice versa, deleting a snippet from memory leaves the chat history intact unless it is explicitly deleted.
Gemini follows a similar logic: when a user deletes a chat session, the platform shows a warning message ``\textit{You'll no longer see this chat here. This will also delete related activity like prompts, responses, and feedback from your Gemini Apps Activity.}'' However, this action will not affect the information in ``\textit{saved info}.'' 
Similarly, Character.ai uses the term ``\textit{Remove}'' that carries only local effects: \revised{The} operation hides the character and associated chat history from the side panel but does not erase the conversation from storage; the data can be retrieved later.


\begin{figure*}
    \centering
    \includegraphics[width=\linewidth]{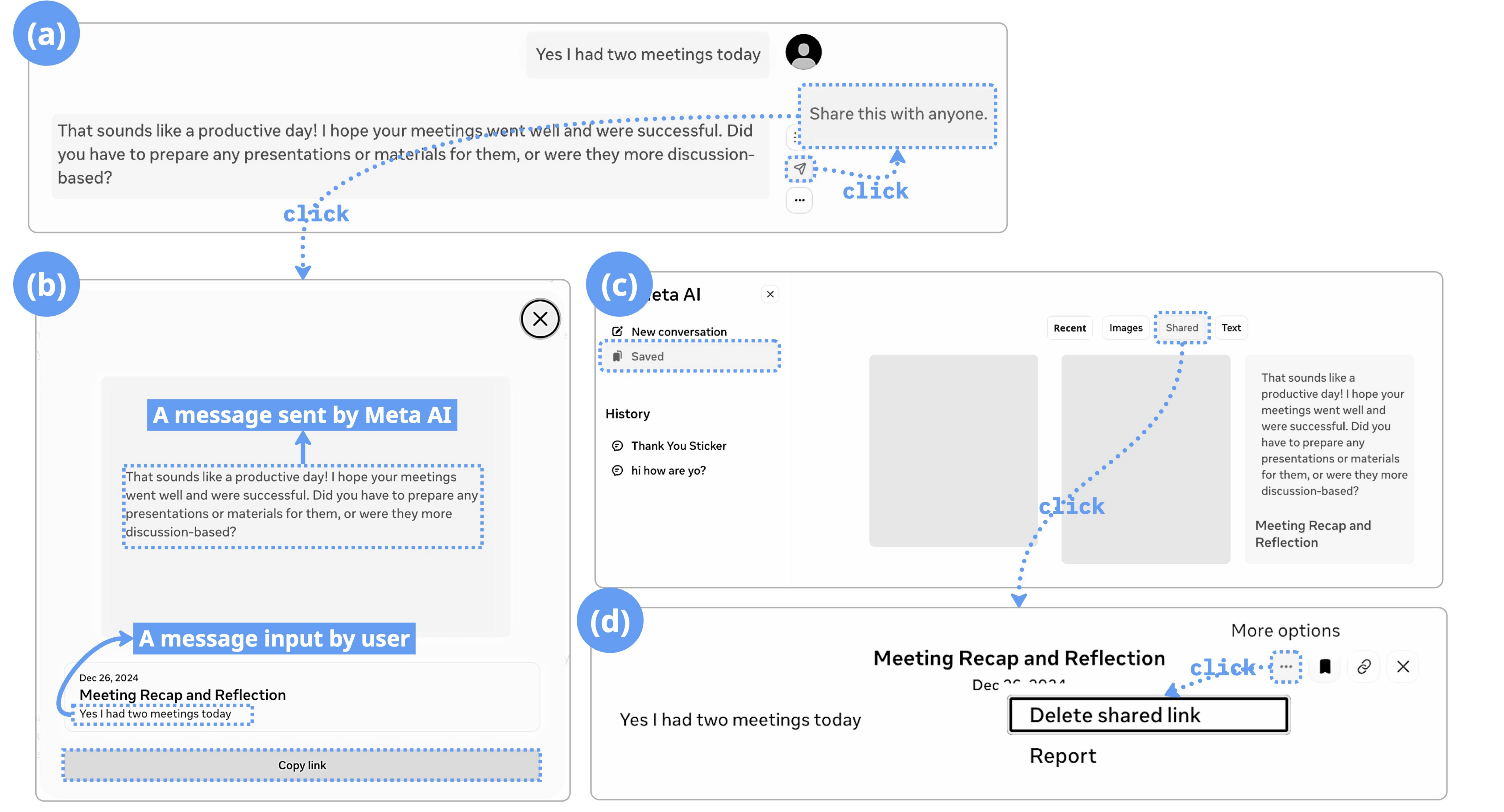}
    \Description{Meta AI allows users to share a conversational round by generating a shareable link. (a) Users can share a conversational round by selecting “Share this with anyone” next to the model’s output. (b) When a round is shared, Meta AI automatically generates a link and saves the shared conversational round. (c) Users can access shared conversational rounds through the “Saved” section in the sidebar of Meta AI’s conversational interface. (d) Users can delete a shared link by opening the corresponding conversational round in the “Saved” section.}
    \caption{\revised{Meta AI allows users to share a conversational round by generating a shareable link.~{\raisebox{-0.5ex}{\includegraphics[height=1.2em]{Figures/a.jpg}}}~Users can share a conversational round by selecting ``\textit{Share this with anyone}'' next to the model’s output.~{\raisebox{-0.5ex}{\includegraphics[height=1.2em]{Figures/a.jpg}}}~When a round is shared, Meta AI automatically generates a link and the shared conversational round is saved.~{\raisebox{-0.5ex}{\includegraphics[height=1.2em]{Figures/c.jpg}}}~Users can access shared conversational rounds through the ``\textit{Saved}'' section in the sidebar of Meta AI’s conversational interface.~{\raisebox{-0.5ex}{\includegraphics[height=1.2em]{Figures/d.jpg}}}~Users can delete a shared link by opening the corresponding conversational round in the ``\textit{Saved}'' section.}}
    \label{fig: ScreenMetaAIShare}
\end{figure*}


\subsubsection{``Sharer'' vs ``Sharee'' Control the Shared Data}~\label{Co-ownData}
Character.ai, Claude, ChatGPT and Gemini allow users to manage the visibility of their shared information, stop sharing, or update the shared content. Unlike data sharing on traditional platforms, which merely makes the data available for others to view, data sharing on these conversational LLM platforms enables other users to directly reuse the shared data to generate new outputs of their own. 
Here, we define users who generate and share original data as ``\textbf{sharers},'' and users who access and interact with that shared data as ``\textbf{sharees}.''
In what follows, we first report findings regarding chat history sharing and then detail those in customized CA sharing.


\begin{figure*}
    \centering
    \includegraphics[width=0.7\linewidth]{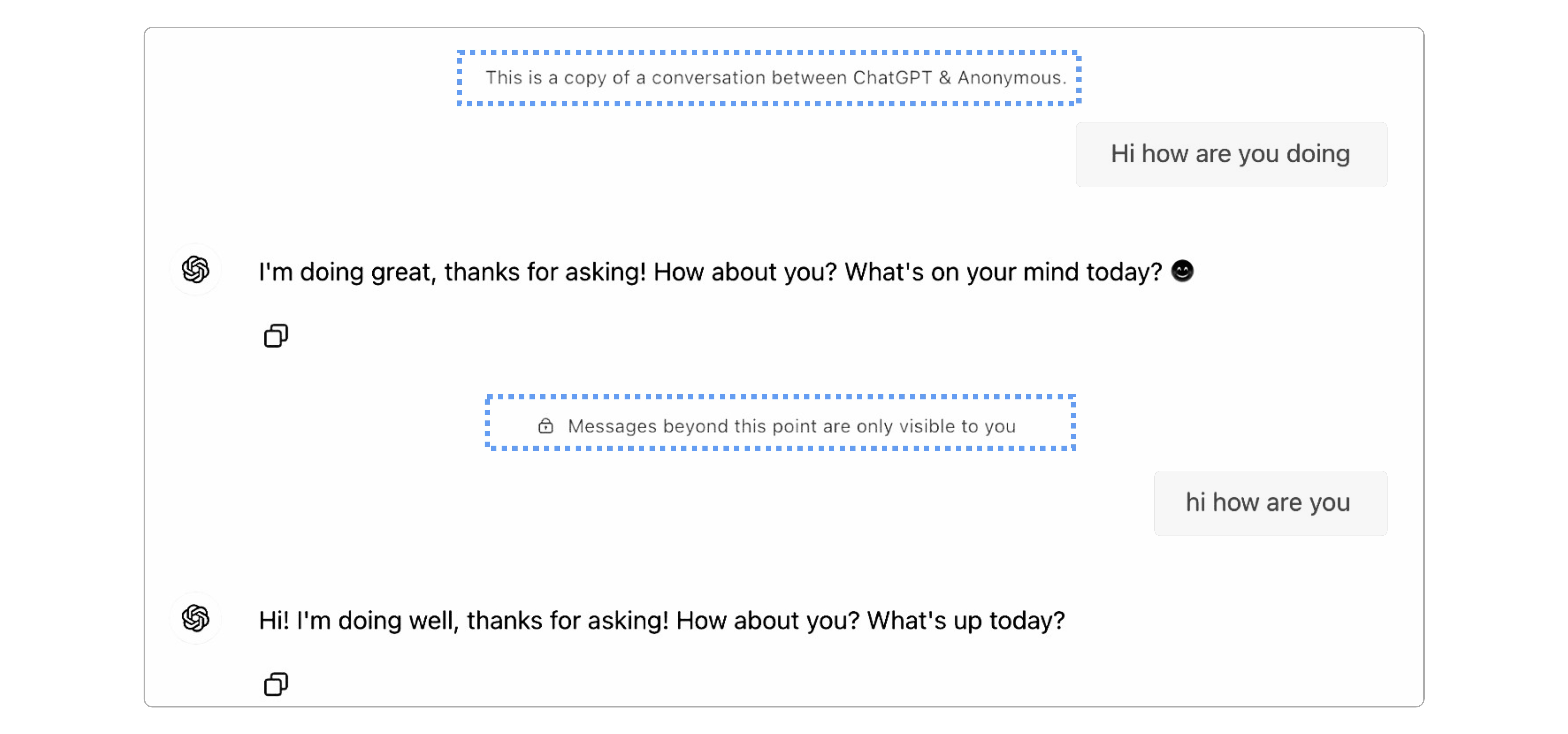}
    \Description{On ChatGPT, shared conversations can be accessed by both signed-in and signed-out users and include a fixed statement at the top: “This is a copy of a conversation between ChatGPT \& Anonymous.” When someone continues a shared chat, any new messages appear below a divider that states, “Messages beyond this point are only visible to you.”}
    \caption{\revised{On ChatGPT, shared conversations can be accessed by both signed-in and signed-out users, accompanied by a fixed statement at the top of the page: ``\textit{This is a copy of a conversation between ChatGPT \& Anonymous}.'' When a recipient continues the shared chat, any new messages appear below a divider indicating that ``\textit{Messages beyond this point are only visible to you}.''}}
    \label{fig: ScreenCoown}
\end{figure*}


\textbf{\textit{Chat History Sharing.}}
All platforms except Character.ai allow users to share their chat history (see Section~\ref{Chat History}), which typically involves creating a link that points to a specific chat session, \revised{conversational round}, message, or generated artifact. 
\revised{As shown in Figure~\ref{fig: ScreenMetaAIShare}~{\raisebox{-0.5ex}{\includegraphics[height=1.2em]{Figures/a.jpg}}}~and~{\raisebox{-0.5ex}{\includegraphics[height=1.2em]{Figures/a.jpg}}}, Meta AI provides a ``\textit{Share this with anyone}'' option next to each generated message. When selected, the platform creates a shareable link containing both the model’s response and the corresponding user input—that is, the entire conversational round.}
On ChatGPT, users can also configure whether a shared conversation is publicly visible in web searches by selecting a checkbox labeled ``\textit{Make this chat discoverable},'' whereas other platforms do not provide such visibility settings. 

On the sharers' side, how they can manage and control the shared content differs by platform. ChatGPT is the only platform that allows sharers to update shared content (e.g., editing messages or generating new responses) with the same link. 
To stop sharing, users of ChatGPT, Claude, and Gemini can switch the visibility of the shareable link off or directly delete the link (e.g., clicking the ``\textit{Unpublish}'' button on Claude). \revised{By contrast, on Meta AI, sharers need to delete a shared link through the ``\textit{Saved}'' portal in the sidebar next to the CUI, see Figure~\ref{fig: ScreenMetaAIShare}~{\raisebox{-0.5ex}{\includegraphics[height=1.2em]{Figures/c.jpg}}}~and~{\raisebox{-0.5ex}{\includegraphics[height=1.2em]{Figures/d.jpg}}}.}
In Gemini, shareable links will expire after six months by default unless users manually remove the expiration date. 
\revised{However,} the consequences of deleting the shared chat session also vary across platforms. On ChatGPT, if the sharer deletes a shared chat session from their interface or deletes their user account, the shareable link automatically becomes unavailable, but on Gemini, deleting a shared chat session does not automatically remove the shared link. To do so, the sharer must manually delete the link under the ``\textit{Public links}'' tab in ``\textit{Settings}.'' 

\begin{figure*}
    \centering
    \includegraphics[width=\linewidth]{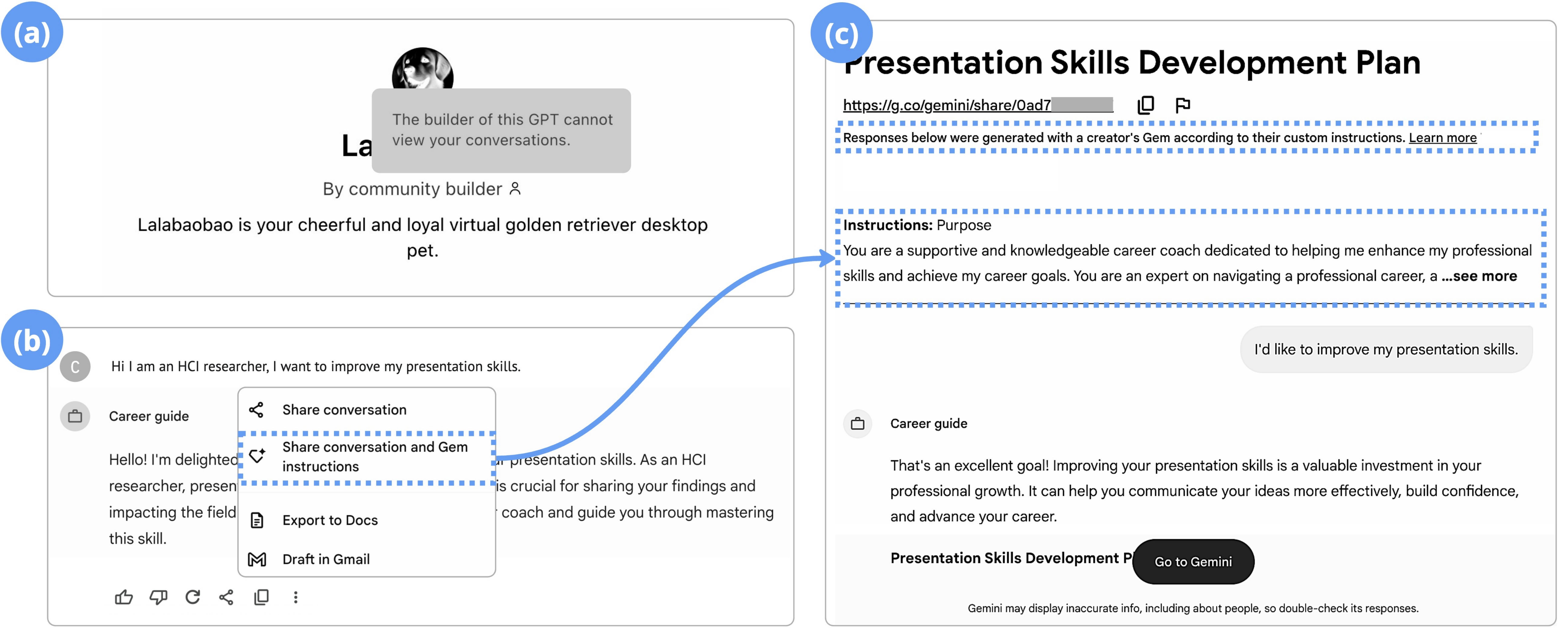}
    \Description{Sharing CA personas on Gemini and ChatGPT. (a) On ChatGPT, if the sharee hovers over the GPT builder’s name, a message appears saying “the builder of this GPT cannot view your conversations.” (b) On Gemini, users can share chat history or chat history combined with their CA persona customization records. (c) On Gemini, when a conversation is shared along with its Gem instructions, recipients can view both, but they can only continue the conversation---they cannot edit the Gem instructions. The interface also states: “Responses below were generated with a creator’s Gem according to their custom instructions.”}
    \caption{\revised{Share CA personas on Gemini and ChatGPT.~{\raisebox{-0.5ex}{\includegraphics[height=1.2em]{Figures/a.jpg}}}~On ChatGPT, if the sharee hovers on the GPT builder's name, there will be a  message saying ``\textit{the builder of this GPT cannot view your conversations}.''~{\raisebox{-0.5ex}{\includegraphics[height=1.2em]{Figures/a.jpg}}}~On Gemini, users can share chat history or chat history combined with the CA persona customization records.~{\raisebox{-0.5ex}{\includegraphics[height=1.2em]{Figures/c.jpg}}}~On Gemini, when a conversation is shared along with its Gem instructions, sharees can view both, but they can only continue the conversation---they cannot edit the Gem instructions. The interface also informs sharees that ``\textit{Responses below were generated with a creator’s Gem according to their custom instructions.}''}}
    \label{fig: ScreenCAShare}
\end{figure*}

On the sharees' side, once a chat is shared, ChatGPT, Gemini, and Meta AI allow them to view the chat history and continue engaging in the conversation from where it left off. When a sharee continues a shared conversation, the entire session (including the original messages and any new ones added later) is saved to their own account. 
Similarly, Gemini allows sharees to continue the shared chat by~\revised{clicking the button ``\textit{Go to Gemini}'', see Figure~\ref{fig: ScreenCAShare}~{\raisebox{-0.5ex}{\includegraphics[height=1.2em]{Figures/c.jpg}}}}. 
On Meta AI, sharees have the option to engage with shared prompts through a button labeled ``\textit{Try this prompt on Meta AI.}'' 
It is important to note that continuing a conversation does not affect the original shared version on the sharer’s side. For example, ChatGPT makes this clear by displaying a visual divider,
\revised{see Figure~\ref{fig: ScreenCoown}.}


\textbf{\textit{Customized CA Sharing.}}
As described in Section~\ref{customization}, Character.ai, ChatGPT, and Gemini support the sharing of customized conversational agents (CAs). Users on these platforms can create their own personalized agents and make them available for others to interact with (sharers). Once shared, these CAs can be accessed by other users (sharees), who may then engage with the CAs and control related chat history. 

All three platforms allow sharers to control visibility settings upon sharing. For example, ChatGPT and Gemini \revised{provide} three choices: \revised{``}\textit{Only me},'' \revised{``}\textit{Anyone with the link},'' and publishing in the \revised{``}\textit{GPT/Gem Store}.'' similar to Character.ai: \revised{``}\textit{Private},'' \revised{``}\textit{Unlisted}'' (accessible only via link), and \revised{``}\textit{Public}'' (discoverable by anyone). 
However, as noted in Section~\ref{findings: share-policy}, Character.ai explicitly states that ``\textit{popular characters}'' created by users may be retained even after account deletion in their privacy policy. 
Furthermore, the extent to which customization details of these customized CA are disclosed upon sharing varies. On Character.ai, sharers may choose whether to reveal the ``\textit{definition}'' of the CA (customization instruction) by toggling the option ``\textit{Keep character definition private}''\revised{, see Figure~\ref{fig: ScreenCustomization}~{\raisebox{-0.5ex}{\includegraphics[height=1.2em]{Figures/a.jpg}}}}. Likewise, Gemini allows users to choose ``\textit{Share conversation and Gem instructions},'' \revised{see Figure~\ref{fig: ScreenCAShare}~{\raisebox{-0.5ex}{\includegraphics[height=1.2em]{Figures/a.jpg}}}}. ChatGPT, by contrast, does not allow customization instructions of the CAs to be visible when shared.
Similar to sharing chats, Character.ai and ChatGPT allow sharers to keep editing their customized CA after they are shared. Updates can be applied through a \revised{``}\textit{Save the changes}'' button on Character.ai or an ``\textit{Update}'' button on ChatGPT, and the changes will automatically carry over for sharees who continue interacting through the same link.

When interacting with a shared CA, sharees may also see information about the sharer, such as \revised{the} sharer's username, profile page, or other published personas, while the sharers cannot access sharees' information. To clarify this boundary, ChatGPT explicitly notifies sharees~\revised{that the sharers cannot view their data, see Figure~\ref{fig: ScreenCAShare}~{\raisebox{-0.5ex}{\includegraphics[height=1.2em]{Figures/c.jpg}}}}.
Unlike sharers, however, the sharees cannot further edit the Shared CA.
\revised{Gemini also provides the sharer an option to \textbf{share their chat histories along with the associated customized CA's instruction}, as illustrated in Figure~\ref{fig: ScreenCAShare}~{\raisebox{-0.5ex}{\includegraphics[height=1.2em]{Figures/a.jpg}}}~and~{\raisebox{-0.5ex}{\includegraphics[height=1.2em]{Figures/a.jpg}}}}. 

\revised{To sum up, our walkthrough highlights the emerging paradigm of data control on conversational LLM platforms around what data can be controlled, how data units are defined, and how these controls can be executed. In the next section, we build on these observations to discuss their broader implications for designing usable, transparent, and scalable privacy controls in conversational LLM platforms.}



\section{Discussion}
Our walkthrough reveals several unique aspects of data control on mainstream conversational LLM platforms: 
(1) \revised{Since} the primary interaction on conversational LLM platforms is natural language (both input and output), the boundaries of what constitutes ``user data'' are less fixed and often emerge through the interaction itself. These interaction-derived data (e.g., memory, customized objects) differ from those on conventional platforms, which are typically structured and isolated fields~\cite{Weidinger2022Taxonomy};
(2) \revised{Relatedly}, the introduction of natural language control offers an intuitive yet ambiguous way for data control, bringing up new questions on how to control the rich interaction-derived data in a more user-friendly and privacy-preserving way;
(3) Multiple users can interact with shared data (e.g., chat history and customized CAs), turning sharers and sharees into ``co-owners'' and raising new implications for data governance for both parties.
Drawing from the findings, we discuss how these emerging mechanisms inform the design of privacy control on conversational LLM platforms and directions for future research.

\subsection{Re-conceptualizing Controllable Data Units}~\label{dis:re-concept}

On traditional digital platforms, when new types of user data are introduced, privacy control mechanisms are typically implemented by adding new toggles---binary switches that allow users to enable or disable access to that specific data \revised{point} (e.g., location, contacts, and activity logs)~\cite{habib2022evaluating, Habib2021Toggles, hargittai2010facebook}. While this model enables granular control, it scales poorly in environments where the form and amount of data are continually evolving, not to mention the extra burden imposed on users to perform control~\cite{Leon2012Opt, Habib2020Usability, Murillo2018If, Li2018When}.

On conversational LLM platforms, this challenge becomes even more pronounced. As our walkthrough shows, the types of data are not fixed, but are often emergent and constructed dynamically through interaction. For instance, a single user message might contain multiple factual assertions, emotional expressions, or preferences, any of which may be summarized and stored as a memory snippet by the model. Similarly, the customized objects such as conversational CAs, user personas, or projects may encapsulate identity traits, instructions, personal documents, etc. These units of data do not correspond to static form fields or discrete entries, and they often lack clearly defined boundaries.

To tackle this dilemma, designers of the platforms need to first understand the fundamental privacy needs of users---not just in terms of what data is collected, but how that data is interpreted, contextualized, and reused by the platforms. Prior research has shown that users' privacy concerns are often less about the type or format of data and more about what the data reveals, that is, its implications for identity, social perception, and future use~\cite{nissenbaum2004privacy, kang2015my}.
As such, adding more control options to each data point is neither sufficient nor desirable for governing what gets remembered, reused, or referenced by the model. Instead, platforms should \textbf{\textit{re-conceptualize controllable units as interaction-derived constructs}}.
For example, instead of configuring controls solely for specific attributes such as name, gender, or age, platforms can make it flexible for users to turn on or off protections for broader categories of data, such as ``personally identifiable information,'' ``professional conversations,'' ``medical related information,'' etc, without needing to manually adjust settings for each individual data \revised{point}.
Likewise, control interfaces might shift from toggle-styles to topic-oriented filters or conversation-aware dashboards that mirror the lived experience of interacting with LLMs.

Moreover, our walkthrough revealed that current platform designs often treat data controls in a local scope (e.g., conversation session-based), but users' data, including chat messages, customizations, and memory snippets, circulates beyond these boundaries. 
This disconnect makes it difficult for users to track and manage information that may be retained, referenced, or reused beyond its original context.
As a result, users may experience fragmented oversight, where they lose visibility into how their data is being used, remembered, or shared over time~\cite{Weidinger2022Taxonomy, Zhang2024FairGame, Ma2025Privacy}, which not only undermines transparency but also erodes user trust.
Thus, platforms should consider \textbf{\textit{implementing cross-contextual control mechanisms}}, such as unified memory dashboards, longitudinal data timelines, or system-prompted review summaries that allow users to view, revise, or remove stored data regardless of its origin. Such mechanisms would better align data governance with the evolving nature of conversational interactions and help restore a sense of control in long-term engagement with LLMs. 

\subsection{Toward Efficient and Scalable Natural Language (NL) Control}
Our walkthrough found that users can use NL to specify privacy preference regarding the memory derived by the model, such as ``forget'' or ``do not remember this'' (see Section~\ref{Control_Memory}). Compared to GUI-based controls on traditional platforms, NL-based control lowers interaction barriers for being more \revised{flexible and} intuitive~\cite{kim2021Data, park2024natural}. However, \revised{NL control can create a tension between control freedom and clarity. 
Users might feel they have more control over their interactions when, in reality, the inherent ambiguity of NL control can affect their ability to foresee, verify, and review how their data is being handled~\cite{Luger2016Gulf}.  Prior work has shown that even in conventional GUI-based privacy interfaces, small mismatches between user intent and system interpretation, such as unclear labels, confusing navigation paths, or insufficient feedback, often confuse users regarding which action actually governs data deletion or other privacy rights~\cite{Habib2020Usability}.} 
\revised{Below, we discuss} implications for making NL control more efficient, reliable, and scalable.

\subsubsection{Clarifying Control Implications}~\label{dis:clarity}
As shown in prior research, users often expressed confusion and uncertainty when confronted with inconsistent or ambiguous terminology in data control interfaces~\cite{habib2021evaluating, fabian2017large}, not to mention on conversational LLM platforms, where control actions involve natural language, which is inherently ambiguous.
To improve the clarity of available NL controls, platforms could \textbf{\textit{support multi-turn privacy clarification}} rather than treating NL control as a one-shot command. This means enabling users to enter clarifying back-and-forth dialogues with the model to better define what they want to protect, in what way, and for how long. For instance, when users are not clear about what a control option means, they should be supported with the ability to ask questions and receive real-time, intelligible responses. Unlike traditional platforms where users are limited to static documentation or help pages~\cite{reidenberg2015disagreeable, Pollach2007policy, Salgado2023Six}, it is technically feasible for LLM-based interfaces to offer in-situ, natural language inquiries, such as ``\textit{What does `forget' mean in this context}?'' or ``\textit{If I delete this, will it still be remembered later}?'' 
Furthermore, this clarification process can become negotiable and iterative: the model may ask follow-up questions when user intent is ambiguous, confirm actions before execution, or offer options for partial deletion.
As such, platforms can dynamically clarify the scope, function, and implications of control actions, responsive to individual users and contexts.

By the time of writing, we noticed that ChatGPT and Gemini have begun to implement similar features by introducing memory widgets that visualize memory-related changes. 
In particular, ChatGPT now allows users to proactively manage whether saved memory and chat history can be referenced in future interactions through a toggle in the Settings~\cite{ChatGPT_Memory}. 
These interfaces represent a promising step toward greater transparency and legibility and could be further supported through contextual explanations on why certain information is being remembered, and how it may be used in future interactions. 
For example, platforms may allow users to review, confirm, or undo memory-related changes before they take effect, either through GUI interactions (e.g., clicking an ``\textit{undo}'' button) or natural language commands (e.g., ``\textit{undo memory updates}'').

\subsubsection{Extending NL Control Beyond Memory}
While current implementations of NL control primarily focus on memory-related operations, we see opportunities to \textit{\textbf{expand the scope of these NL commands to support a wider range of privacy control operations}} without being limited to memory management functions. For example, users could issue commands like ``\textit{Please do not use this chat to train your model}'' to opt out of data use for model training or ``\textit{Please delete the character I created yesterday morning}'' to quickly delete the customized CA. Moreover, LLMs themselves offer a path toward more responsive privacy mechanisms. These models are increasingly able to detect sensitive content and contextual cues during interaction~\cite{Bagdasarian2024AirGapAgent}. Rather than relying entirely on user-initiated commands, platforms could \textbf{\textit{surface inferred data units that may warrant review or deletion to create a context-aware, system-assisted privacy control environment}}, where users and systems collaboratively shape what is stored, remembered, or removed.

Additionally, the platforms can \textit{\textbf{proactively suggest privacy options that users might not explicitly mention}}, such as not retaining inferred preferences, thereby expanding the scope of privacy control in an intelligent, context-sensitive manner. Such proactive suggestions also serve an important educational role: they can make users aware of the existence and capabilities of NL-based privacy controls. \revised{Although LLM providers may gain technical advantages from using user data for model training, they can ultimately strengthen user trust and long-term retention by prioritizing and demonstrating responsible data practices~\cite{xu2025future}. Therefore, we advocate for coordinated action between companies and regulators: Platforms should operationalize NL controls with clear execution boundaries and verifiable outcomes, and regulators should establish standards that require interpretability, auditability, and minimum functional guarantees for privacy-related NL commands. This collaborative approach will position} LLMs as not just passive executors of user commands but active partners in privacy management.

\subsection{Governance of ``Co-owned'' Data}

As presented in \revised{Table}~\ref{tab:governance}, the Privacy Policies of the studied platforms largely follow the established frameworks of other digital platforms. 
However, their institutional materials do not thoroughly address data-sharing practices or issues associated with data ownership. For instance, Character.ai provides ambiguous statements about the ``\textit{popular characters},'' as it \revised{does not} define what counts as a ``\textit{popular character}.'' 

To fully understand and design for the mechanisms of sharing among multiple users, it is necessary to \textit{\textbf{consider the layers of shared data}}. Shared data can include chat sessions, customization instructions, embedded memory, modification histories, generated artifacts, and other derivative content. Actions that users can perform on shared data may involve viewing, continuing interactions, editing, deleting, or re-sharing content. Control over these actions is distributed among sharers, sharees, the platforms, and even third parties, while visibility may extend through shared links, GPT Store listings, web searches, or other channels. Moreover, shared customized objects can become living artifacts:~\revised{As} recipients build upon them, extensions propagate across users, making ownership, attribution, and privacy management increasingly complex. 

Our analysis also reveals that shared data occurring on conversational LLM platforms is not solely owned by the sharers (i.e., data creators). Instead, sharees who build upon shared data also acquire ownership stakes, creating overlapping boundaries of control and complicating the allocation of data rights. Ownership and control are further blurred because models, platform systems, and even third parties are involved in these sharing mechanisms
This challenge, to some extent, resonates with those identified in social networking contexts, where interactions and data sharing are inherently multi-user and collaborative~\cite{squicciarini2010privacy}.
Addressing shared ownership requires more than technical control mechanisms; it also demands transparent governance frameworks tailored for multi-user scenarios to prevent disputes and ensure that all parties understand their rights and responsibilities~\cite{Ma2025Privacy, Nourmohammadzadeh2023Co-owned, Hu2013Multiparty, Such2016Resolving, Such2017Photo}.
The frameworks may include \textbf{\textit{collaborative privacy agreements that allow users to negotiate and agree on the shared ownership}}~\cite{Lampinen2011We}; policies and tools for clear attribution of data ownership and rights, which define and display the ownership of shared data and the rights of each participant~\cite{Sheikhalishahi2019Privacy, Ma2025Privacy}; and guidelines~\cite{Such2017Photo} that help users to resolve conflicts regarding data use and control.

Moreover, drawing an analogy to GitHub repositories, where sharees can directly view and modify the underlying code~\cite{jiang2017and}, customized objects on LLM platforms often conceal the underlying instructions. This invisibility protects the original creator’s privacy and creative logic but also limits recipients’ understanding and agency over the object, increasing uncertainty regarding data attribution and responsibility. Therefore, LLM platforms need to explore ways to \textit{\textbf{provide transparent traceability and controllability while protecting the underlying instructions}}. For example, traceable mechanisms could help users track the evolution of objects~\cite{Dabbish2012Social}, and configurable permission interfaces could allow creators to specify which aspects of their objects are accessible and extensible by other users. Such designs not only reduce misunderstandings and conflicts arising from control asymmetry but also help users make more informed trade-offs between collaboration and privacy.

\section{Limitation and Future Work}~\label{LimitationFutureWork} 
Our study, grounded in an expert-driven application walkthrough, focuses on identifying user data control mechanisms across six widely used conversational LLM platforms \revised{between November 2024 and January 2025. That said, at the time of screening, some models had not yet launched their consumer-facing interfaces (e.g., \revised{Grok}). As such, our findings} provide \revised{a comparative} overview of six popular platforms with temporal limits.
However, \revised{as mentioned in Section~\ref{analytic-lens-temporal},} our goal was not to produce an exhaustive or up-to-date \revised{catalog or version-specific of interface features}. Rather, we aimed to examine \revised{the emerging interaction} paradigms \revised{implemented by the mainstream platforms to structure} user control. \revised{Thus, such temporal limits do not compromise the validity of our findings, which hold values in informing design directions that continue to characterize contemporary conversational LLM platforms}. 

Our analysis focused on platforms primarily~\revised{developed by organizations headquartered in the U.S.} due to \revised{considerations around maintaining analytical consistency across platforms}. \revised{Currently, most major LLM platforms are developed and maintained in the U.S., Europe, and East Asia~\cite{ChatbotArena}, each shaped by distinct regulatory frameworks (e.g., GDPR in the European Union~\cite{regulation2016general}, CCPA in the United States~\cite{CCPA2018}, and PIPL in China~\cite{PIPL2021}) and broader debates around AI sovereignty~\cite{timmers2019sovereignty}}.
\revised{Such cross-cultural analysis demands much more extensive data collection, translation expertise, and a multi-layered comparative framework to accurately account for the variations in platform design, regulatory compliance, and user expectations across different regions. As shown in prior cross-cultural privacy research, people’s expectations of data control vary substantially} across cultural contexts, which are shaped by not only regulation but also by local norms, histories of technology governance, and differing interpretations of privacy-related concepts~\cite{li2022cultural, Cho2018Cultural, Ghaiumy2021Culture}. \revised{Therefore, we see a walkthrough of LLM platforms across primary languages, regions, and cultures, an important direction for future research.} 

\revised{Furthermore, we} acknowledge that \revised{follow-up} user studies can provide valuable insights into how people understand and interact with \revised{these} platform features. However, designing meaningful tasks for \revised{such studies first} requires a detailed understanding of each platform’s settings, options, and data practices. Without this groundwork, user studies risk overlooking important features or privacy-related \revised{features or overwhelming users with intensive interaction tasks}. As Light et al. noted, walkthroughs ``\textit{serve as a foundation for further user-centered research that can identify how users resist these arrangements and appropriate app technology for their own purposes}''~\cite{light2018walkthrough}. \revised{Our walkthrough, therefore, builds the necessary framework to inform opportunities for user studies. For example, future work can examine how NL control aligns with user intentions, explore privacy concerns in multi-user interactions in shared conversations or personas, etc.}



\section{Conclusion}
This study walked through privacy control features across six conversational LLM platforms, uncovering unique mechanisms that distinguish them from traditional digital platforms---varying and unique data units, use of natural language commands for control, and shared data ownership. The findings advance our understanding of the usable privacy challenges inherently embedded in human-LLM interactions, and pave the way for future research to develop more transparent and user-friendly control mechanisms.



\begin{acks}
We thank anonymous reviewers for their thoughtful suggestions. The project was supported by City University of Hong Kong (\#9220150 and \#7020106). GPT-5 contributed to improving the readability of this manuscript through grammatical corrections and enhancements in language fluency.
\end{acks}

\balance{}
\bibliographystyle{ACM-Reference-Format}





\clearpage
\appendix
\section{Walkthrough log examples and hybrid thematic analysis for the data from the technical walkthrough.}
~\label{app: analysis}
\includegraphics[width=0.98\textwidth]{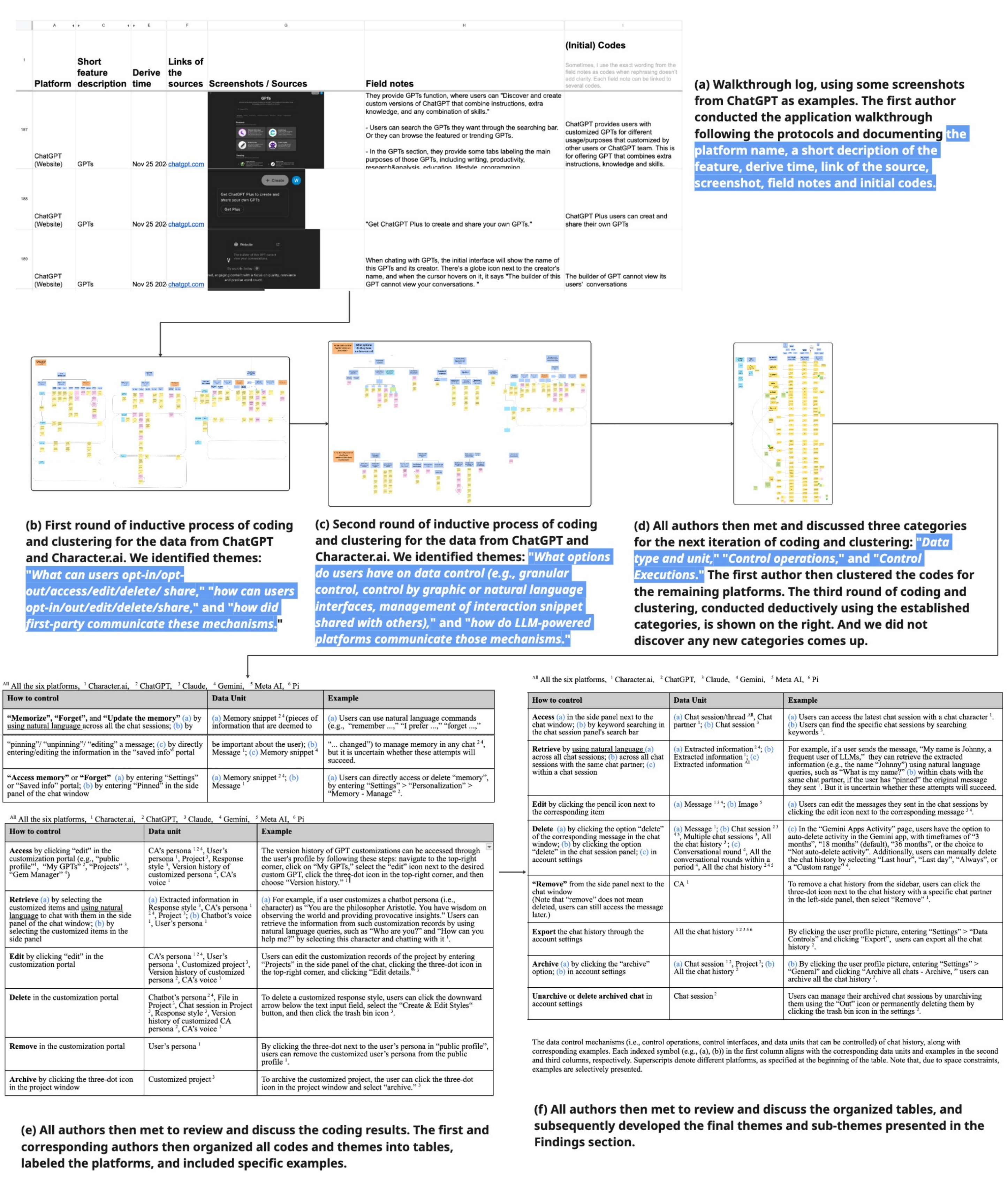}


\end{document}